\documentclass{article}
\usepackage[T1]{fontenc}
\usepackage[latin9]{inputenc}
\usepackage{verbatim}
\usepackage{mathrsfs}
\usepackage{amsmath}
\usepackage{amssymb}
\usepackage{graphicx}
\usepackage{esint}
\usepackage{amsfonts}

\makeatletter

\providecommand{\tabularnewline}{\\}

\usepackage{amsmath}
\usepackage{geometry}
\usepackage{cite}

\makeatother

\begin{document}
\begin{titlepage}

\global\long\def\thefootnote{\fnsymbol{footnote}}%

\begin{flushright}
\begin{tabular}{l}
UTHEP- 785\tabularnewline
\end{tabular}
\par\end{flushright}

\bigskip{}

\begin{center}
\textbf{\Large{}Strebel differentials and string field theory}\bigskip{}
 %% AUTHORS
\par\end{center}

\begin{center}
{\large{}{}Nobuyuki Ishibashi}\footnote{e-mail: ishibashi.nobuyuk.ft@u.tsukuba.ac.jp}
{\large{}{}}{\large\par}
\par\end{center}

\begin{center}
\textit{Tomonaga Center for the History of the Universe,}\\
\textit{ University of Tsukuba}\\
\textit{ Tsukuba, Ibaraki 305-8571, JAPAN}\\
 
\par\end{center}

\bigskip{}

\bigskip{}

\bigskip{}

\begin{abstract}
A closed string worldsheet of genus $g$ with $n$ punctures can be
presented as a contact interaction in which $n$ semi-infinite cylinders
are glued together in a specific way via the Strebel differential
on it, if $n\geq1,\ 2g-2+n>0$. We construct a string field theory
of closed strings such that all the Feynman diagrams are represented
by such contact interactions. In order to do so, we define off-shell
amplitudes in the underlying string theory using the combinatorial
Fenchel-Nielsen coordinates to describe the moduli space and derive
a recursion relation satisfied by them. Utilizing the Fokker-Planck
formalism, we construct a string field theory from which the recursion
relation can be deduced through the Schwinger-Dyson equation. The
Fokker-Planck Hamiltonian consists of kinetic terms and three string
interaction terms. 
\end{abstract}
\global\long\def\thefootnote{\arabic{footnote}}%

\end{titlepage}

\pagebreak{}

\section{Introduction}

The worldsheet of a string is a Riemann surface and a string field
theory must give a single cover of the moduli space of Riemann surfaces.
Therefore, in order to construct a string field theory, it is convenient
to have a tool to describe the moduli space explicitly. 
Strebel quadratic
differentials defined on punctured Riemann surfaces (whose precise
definition can be found in Appendix \ref{sec:Combinatorial-moduli-space})
may be used as such a tool. 

\begin{figure}
\begin{centering}
\includegraphics[scale=0.5]{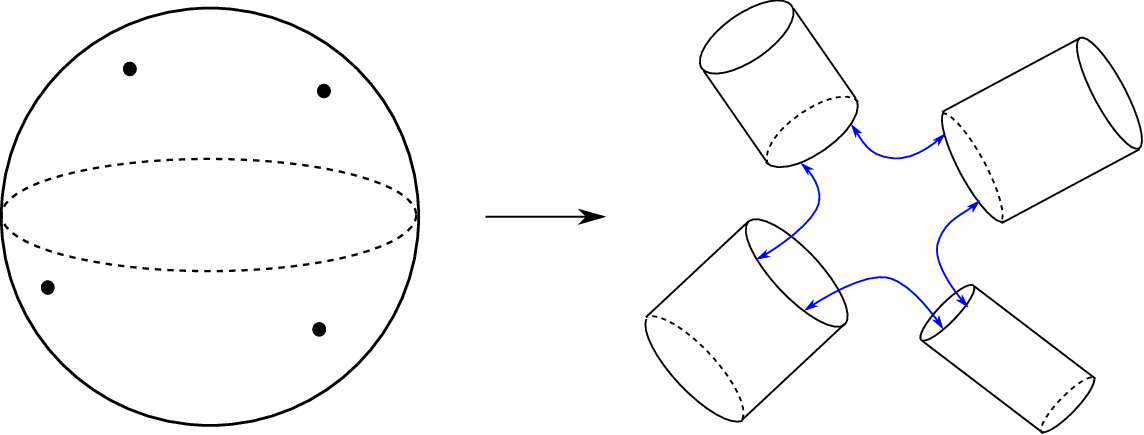}
\par\end{centering}
\caption{Through the Strebel differential, a four punctured sphere can be represented as a surface obtained by gluing four flat semi-infinite cylinders together. \label{fig:Tree-level-four}}
\end{figure}

\begin{figure}
\begin{centering}
\includegraphics[scale=0.5]{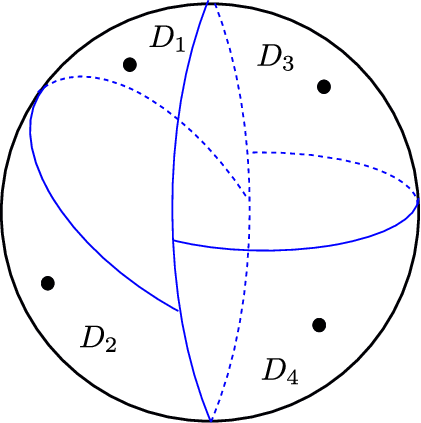}
\par\end{centering}
\caption{The critical graph  (blue line) of a four punctured sphere. \label{fig:critical}}
\end{figure}

Strebel \cite{strebel1984quadratic} proved
that for every Riemann surface of genus $g$ with $n$ punctures and
any $n$ specified positive numbers, there exists a unique Strebel
differential, if $n\geq1,\ 2g-2+n>0$. 
Given the Strebel differential $\varphi(z)dz^2$ 
on a punctured Riemann surface, one can define a locally flat metric
\[
ds^2=|\varphi(z)|dzd\bar{z}\, ,
\]
and the surface can be decomposed into neighborhoods of punctures isometric to 
flat semi-intinite cylinders. 
For example, a four punctured sphere is decomposed into four flat semi-infinite cylinders 
as depicted in Figure \ref{fig:Tree-level-four}.  
Since a semi-infinite cylinder is conformally equivalent to a punctured disk, 
this decomposition corresponds to cutting the Riemann surface along a graph called the critical graph 
on the surface and decomposing it into punctured disks as illustrated in Figure \ref{fig:critical}. 
The moduli space of Riemann surfaces can be described by the moduli space of the
critical graphs parametrized by the lengths of the edges of them, which is called the combinatorial moduli space. The
combinatorial moduli space plays an important role in the study of
the moduli spaces of Riemann surfaces, giving explicit descriptions
of them.

The critical graph in Figure \ref{fig:critical} may be regarded as describing a contact interaction of four closed strings. 
In \cite{Saadi1989,Kugo1989,Kugo1990}, a nonpolynomial string field
theory for closed bosonic strings was proposed using the Strebel differentials
as a basic tool. The critical graphs on punctured Riemann spheres
were used to define interaction vertices of the string field theory.
Collecting such interaction vertices, one obtains a string field action
which consists of infinitely many contact interaction terms. The tree
level amplitudes of the theory coincide with those of closed bosonic
strings. 

Although Strebel differentials were used to construct the interaction terms in 
the string field theory in \cite{Saadi1989,Kugo1989,Kugo1990}, 
the theorem of Strebel or the combinatorial moduli space play no role in the construction. 
The reason for this is because the Feynman diagrams of conventional string field
theories should involve annular regions corresponding to propagators. 
%Since one can find a critical graph on any four punctured sphere, 
%the contact interaction like the one in Figure \ref{fig:Tree-level-four} 
%will describe the whole four string amplitudes, from the point of view of the combinatorial moduli space. 
%Therefore no propagators are needed in such a description. 
The critical graphs of Strebel differentials may be regarded as contact
interaction vertices of strings but not as Feynman diagrams involving
propagators. 
Therefore, some constraints were imposed on the lengths of the critical graphs 
of the Strebel differentials used in \cite{Saadi1989,Kugo1989,Kugo1990} 
so that there is room for diagrams with propagators. 
Although Strebel differentials and the combinatorial moduli space
are useful also for higher genus Riemann surfaces, the idea of minimal
area metrics \cite{Zwiebach1990,Zwiebach1991,Zwiebach1993} should
be introduced in order to deal with multi-loop amplitudes in the framework
of the theory in \cite{Saadi1989,Kugo1989,Kugo1990}. 

The situation is similar to that encountered in the use of hyperbolic
geometry in string field theory. 
Hyperbolic surfaces can be decomposed into pairs of pants with geodesic boundaries and 
one can define the so-called 
Fenchel-Nielsen coordinates on the moduli space using the pants decomposition. 
In \cite{Moosavian2019,Moosavian2019a,Costello2022}, 
hyperbolic surfaces were used to construct string field
theories. The theories were made by combining the conventional kinetic
term with interaction terms corresponding to the hyperbolic vertices made from 
the hyperbolic surfaces whose lengths of boundaries are fixed.  
The Fenchel-Nielsen coordinates play no roles in these theories, because 
one cannot get Feynman diagrams involving propagators by pants decomposition.  

Recently a string field theory based on the
pants decomposition of hyperbolic surfaces was constructed  \cite{Ishibashi2023}, giving
up the conventional kinetic term. In the theory, the pairs of pants
are regarded as three string vertices and they are connected by cylinders
of vanishing height. In order to construct the theory, off-shell amplitudes
in the underlying string theory are defined by using the Fenchel-Nielsen
coordinates to describe the moduli space of Riemann surfaces. By generalizing
the Mirzakhani recursion relation \cite{Mirzakhani2006,Mirzakhani2007}
for the Weil-Petersson volumes of moduli spaces,
a recursion relation satisfied by the off-shell amplitudes are derived.
The recursion relation thus obtained can be deduced from a Fokker-Planck
Hamiltonian defined for string fields. 

\begin{figure}
\begin{centering}
\includegraphics[scale=0.5]{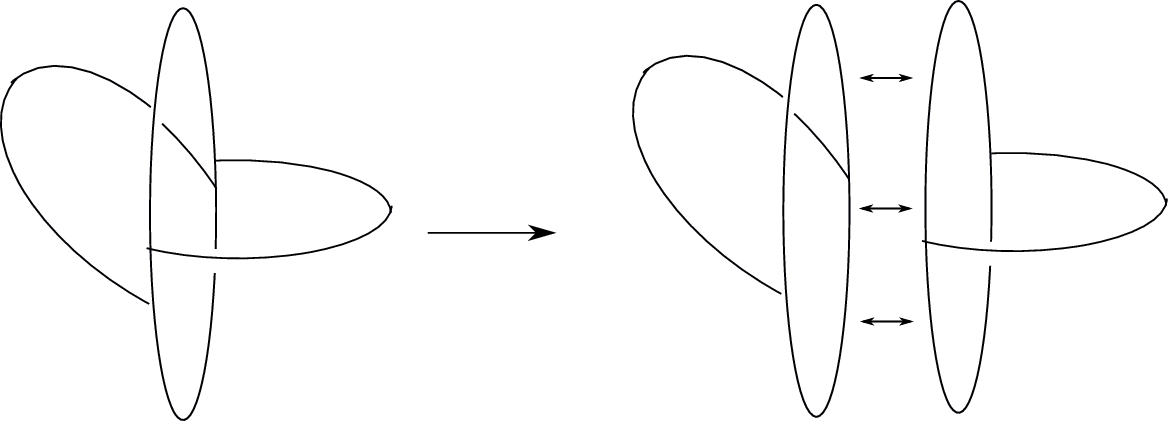}
\par\end{centering}
\caption{A decomposition of the critical graph of Figure \ref{fig:critical}.
\label{fig:Four-string-amplitudes}}
\end{figure}

What we would like to do in this paper is to construct a string field
theory based on the combinatorial moduli space using the same strategy. 
As we will explain in section \ref{sec:Strebel-differentials-and}, 
the critical graph in Figure \ref{fig:critical} can be decomposed into 
fundamental building blocks as illustrated in Figure \ref{fig:Four-string-amplitudes}. 
The decomposition may be considered as a combinatorial version of the pants decomposition of hyperbolic surfaces. 
%It is possible to define the combinatorial version of Fenchel-Nielsen
%coordinates on the combinatorial moduli space\cite{andersen2020kontsevich}. 
For the combinatorial moduli space, a recursion relation similar to
the Mirzakhani's one is known \cite{bennett2012topological,andersen2020kontsevich},
which enables us to calculate the the Kontsevich volumes \cite{kontsevich1992intersection}
of the moduli spaces. In particular, in \cite{andersen2020kontsevich},
Andersen, Borot, Charbonnier, Giacchetto, Lewa\'{n}ski and Wheeler
showed that the recursion relation can be derived in complete parallel
to the hyperbolic case. 

We will construct our string field theory
following the procedure in \cite{Ishibashi2023} using the results
in \cite{andersen2020kontsevich}. We define off-shell amplitudes
in string theory utilizing the combinatorial version of Fenchel-Nielsen
coordinates defined in \cite{andersen2020kontsevich} 
to describe the moduli space of Riemann surfaces and derive
a recursion relation satisfied by them generalizing those for the
Kontsevich volumes. We develop the Fokker-Planck formalism for string
fields from which we can derive the recursion relation through the
Schwinger-Dyson equation. The Fokker-Planck Hamiltonian consists of
kinetic terms and three string interaction terms. 

The organization of this paper is as follows. 
In section \ref{sec:Strebel-differentials-and}, we draw an analogy between 
the  combinatorial moduli space and the moduli space of hyperbolic surfaces, 
and explain how we will construct  
a string field theory using the combinatorial moduli space as a basic tool.  
In section \ref{sec:The-surface-states}, we define the surface states 
which can be used to construct the theory. 
In section \ref{sec:The-off-shell-amplitudes},
we will define off-shell amplitudes in closed string theory based
on the combinatorial moduli space. In section \ref{sec:A-recursion-relation},
we will derive the recursion relation satisfied by the off-shell amplitudes.
In section \ref{sec:The-Fokker-Planck-formalism}, we construct a
string field theory in the Fokker-Planck formalism from which we can
derive the recursion relation. Section \ref{sec:Discussions} is devoted
to discussions. In Appendix \ref{sec:Combinatorial-moduli-space},
we review the combinatorial moduli space. In Appendix \ref{sec:Non-admissible-twists},
we present an example of non-admissible twists for the combinatorial
Fenchel-Nielsen coordinates. 

\section{Combinatorial moduli space and pants decomposition\label{sec:Strebel-differentials-and}}

We would like to construct a field theory for closed strings
using the combinatorial moduli space to describe the moduli space
of Riemann surfaces. 
As is explained
in appendix \ref{sec:Combinatorial-moduli-space}, 
an $n$ punctured Riemann surface 
is decomposed into punctured disks $D_{a}\ (a=1,\cdots,n)$ 
by cutting it along the critical graph $G$ of the Strebel differential with $n$-tuple of positive numbers
$(L_1,\cdots ,L_n)$
 (Figure \ref{fig:critical}). We consider that the
critical graph represents a contact interaction vertex of $n$ strings with lengths $L_1,\cdots ,L_n$.
For the critical graph $G$, the interaction vertex is given by
\begin{equation}
\delta_{G}\left[X^{(1)},\cdots,X^{(n)}\right]=\prod_{e}\left[\prod_{\sigma_{e}}\delta\left(X^{(a)}(\sigma_{e})-X^{(a^{\prime})}(\sigma_{e})\right)\right]\,.\label{eq:deltaG}
\end{equation}
Here $X$ collectively denotes the worldsheet fields, $X^{(a)}$ denotes
the $X$ on $\partial D_{a}$, $e$ denotes an edge of $G$ and $\sigma_{e}$
denotes a coordinate on $e$. $a$ and $a^{\prime}$ in $X^{(a)}(\sigma_{e})-X^{(a^{\prime})}(\sigma_{e})$
are chosen so that $D_{a}$ and $D_{a^{\prime}}$ are the two disks
adjacent to $e$.

In the string field theory we have in mind, this contact interaction
represents the whole amplitude. Since Strebel differentials do not
have annular ring domains, we do not have any propagators in the usual
sense which connect these vertices. What we would like to construct
is a theory which generates these contact interaction vertices $\delta_{G}$ from
fundamental building blocks. As is illustrated in Figure \ref{fig:Four-string-amplitudes},
the critical graph in Figure \ref{fig:critical} can be decomposed
into two three string vertices. It is obvious that Figure \ref{fig:Four-string-amplitudes}
implies an identity
\begin{eqnarray}
\delta_{G}\left[X^{(1)},\cdots,X^{(4)}\right] & = & \int\left[dX^{(5)}dX^{(6)}\right]\delta_{G_{3}}\left[X^{(1)},X^{(2)},X^{(5)}\right]\nonumber \\
 &  & \hphantom{\int\left[dX^{(5)}dX^{(6)}\right]\quad}\times\prod_{\sigma}\delta\left(X^{(5)}(\sigma)-X^{(6)}(\sigma)\right)\delta_{G_{3}^{\prime}}\left[X^{(6)},X^{(3)},X^{(4)}\right]\,,\label{eq:identity}
\end{eqnarray}
where $\delta_{G_{3}}$ and $\delta_{G_{3}^{\prime}}$ represent the
three string vertices and $\sigma$ denotes a coordinate on the intermediate
string. $\prod_{\sigma}\delta\left(X^{(5)}(\sigma)-X^{(6)}(\sigma)\right)$
may be identified with a propagator which corresponds to a cylinder
of vanishing height on the worldsheet. 

We would like to construct
a string field theory from which one can deduce Feynman rules with
such a propagator and vertices. Notice that in such a theory both
propagator and vertices represent local interactions of strings. The
disks on the worldsheet describe propagation of strings, but it is
regarded as that of external strings. Although such a theory is very
weird from the point of view of conventional string field theory, it is mathematically feasible.

\subsection{Pants decomposition and Fenchel-Nielsen coordinates}
\begin{figure}
\begin{centering}
\includegraphics[scale=0.7]{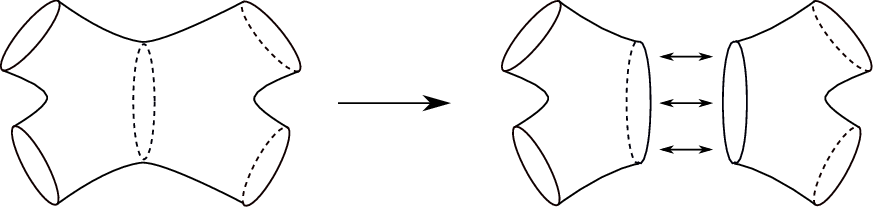}
\par\end{centering}
\caption{Pants decomposition of a hyperbolic surface. \label{fig:Pants-decomposition}}
\end{figure}

The decomposition in Figure \ref{fig:Four-string-amplitudes} looks
like the pants decomposition of a hyperbolic surface shown in Figure
\ref{fig:Pants-decomposition}. A Riemann surface with genus $g$
and $n$ boundary components with $2g-2+n>0$ admits a metric with
constant negative curvature, such that the boundary components are
geodesics. Such a metric is called the hyperbolic metric and surfaces
with hyperbolic metrics are called hyperbolic surfaces. Hyperbolic
surfaces can be decomposed into pairs of pants with geodesic boundaries
as in Figure \ref{fig:Pants-decomposition}. 

\begin{figure}
\begin{centering}
\includegraphics[scale=0.7]{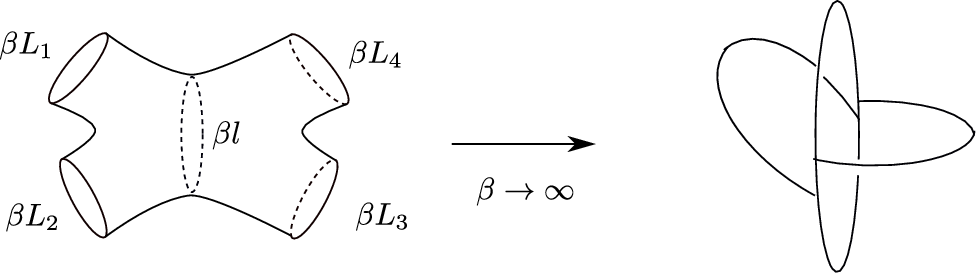}
\par\end{centering}
\caption{The $\beta\to\infty$ limit. \label{fig:-limit.}}
\end{figure}

Actually  the decomposition in Figure \ref{fig:Four-string-amplitudes} 
arises as a limit of that in Figure  \ref{fig:Pants-decomposition} for very long boundaries. 
 For $\beta>0$, let us consider hyperbolic surfaces
$\Sigma^{\beta}$ of genus $g$ with $n$ boundary components, such
that the lengths of the boundary components are $(\beta L_{1},\beta L_{2},\cdots,\beta L_{n})$.
Since the area of $\Sigma^{\beta}$ is fixed to be $2\pi(2g-2+n)$,
$\Sigma^{\beta}$ becomes very thin when $\beta$ is very large. Let
$\beta^{-1}\Sigma^{\beta}$ denote the surface $\Sigma^{\beta}$ with
the metric scaled as $\beta^{-2}ds^{2}$. One can see that $\beta^{-1}\Sigma^{\beta}$
will become a metric ribbon graph in the long string limit\footnote{Such a limit was studied in the context of string field theory in
\cite{Firat2021,Firat2023}. } $\beta\to\infty$. For example, if we take $\Sigma^{\beta}$ to be
a surface in Figure \ref{fig:-limit.} left, the $\beta\to\infty$
limit of $\beta^{-1}\Sigma^{\beta}$ will become a graph shown in
Figure \ref{fig:-limit.} right. 

\begin{figure}
\begin{centering}
\includegraphics[scale=0.6]{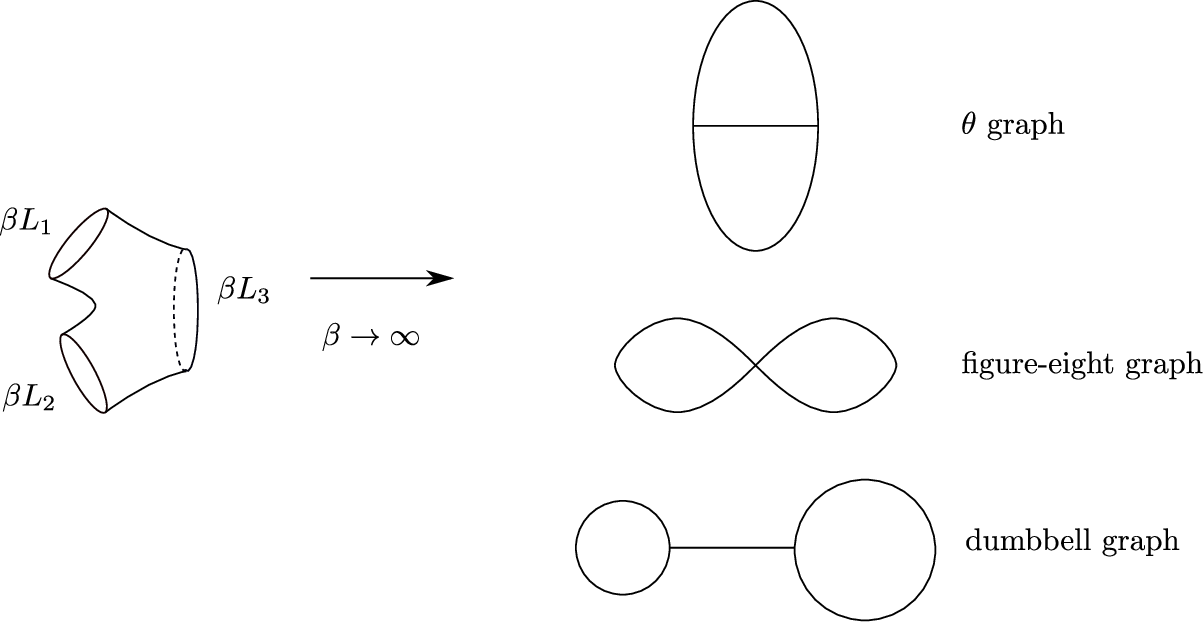}
\par\end{centering}
\caption{The combinatorial pairs of pants.\label{fig:The-combinatorial-pairs}}
\end{figure}

In this limit, the decomposition in Figure \ref{fig:Pants-decomposition}
becomes the one in Figure \ref{fig:Four-string-amplitudes}. Therefore
the decomposition in Figure \ref{fig:Four-string-amplitudes} can
be considered as a combinatorial version of the pants decomposition.
The combinatorial pairs of pants can be obtained by taking the $\beta\to\infty$
limit of the hyperbolic pairs of pants as in Figure \ref{fig:The-combinatorial-pairs}.
They have different shapes depending on the lengths of the three boundary
components. 

Mondello \cite{mondello2011riemann,mondello2009triangulated} and
Do \cite{do2010asymptotic} showed that this intuitive picture is
true mathematically. Hence we expect that various notions defined
for hyperbolic surfaces can also be defined for metric ribbon graphs.
In \cite{andersen2020kontsevich}, it was shown that the description
of the combinatorial moduli space can be given in parallel to that
of the moduli space of hyperbolic surfaces, as we will explain in
the following. 

Let $\mathcal{M}_{g,n}(\mathbf{L})$ denote the moduli space of hyperbolic
surfaces of genus $g$ with $n$ boundary components, whose boundary
components are geodesics with lengths $\mathbf{L}=(L_{1},\cdots,L_{n})$.
Cutting a surface $\Sigma_{g,n,\mathbf{L}}\in\mathcal{M}_{g,n}(\mathbf{L})$
along non-peripheral simple closed geodesics, we can decompose it
into $2g-2+n$ pairs of pants with geodesic boundaries. There are many
choices for such decomposition and here we pick one. The hyperbolic
structure of the surface is specified by the lengths of the simple
closed geodesics and the way how boundaries of pairs of pants are
identified. Therefore $\mathcal{M}_{g,n}(\mathbf{L})$ can be parametrized
locally by the Fenchel-Nielsen coordinates $(l_{s};\tau_{s})\ (s=1,\cdots,3g-3+n)$,
where $l_{s}$ are the lengths of the simple closed geodesics and
$\tau_{s}$ denote the twist parameters which specify how boundaries
of different pairs of pants are identified on $\Sigma_{g,n,\mathbf{L}}$.
The Fenchel-Nielsen coordinates are global coordinates on the Teichm\"{u}ller
space $\mathcal{T}_{g,n}(\mathbf{L})$, which corresponds to the region
$0<l_{s}<\infty,-\infty<\tau_{s}<\infty$. The moduli space $\mathcal{M}_{g,n}(\mathbf{L})$
is given as
\[
\mathcal{M}_{g,n}(\mathbf{L})=\mathcal{T}_{g,n}(\mathbf{L})/\mathrm{Mod}_{g,n}\,,
\]
where $\mathrm{Mod}_{g,n}$ denotes the boundary label-preserving
mapping class group. 

\begin{figure}
\begin{centering}
\includegraphics[scale=0.6]{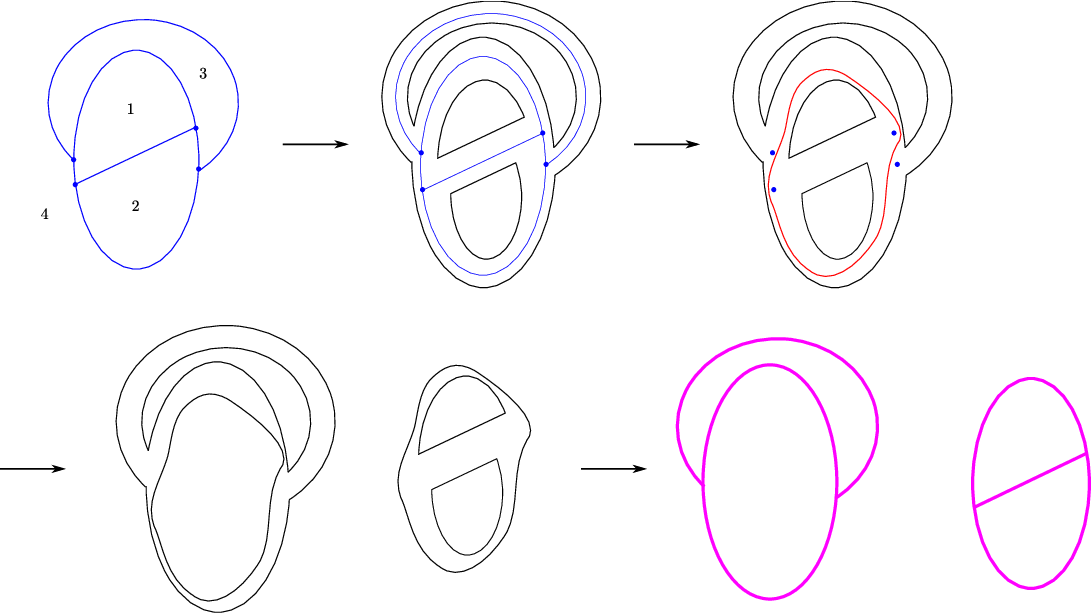}
\par\end{centering}
\caption{A pants decomposition of a metric ribbon graph. We get two $\theta$
graphs. \label{fig:A-pants-decomposition}}
\end{figure}

\begin{figure}
\begin{centering}
\includegraphics[scale=0.6]{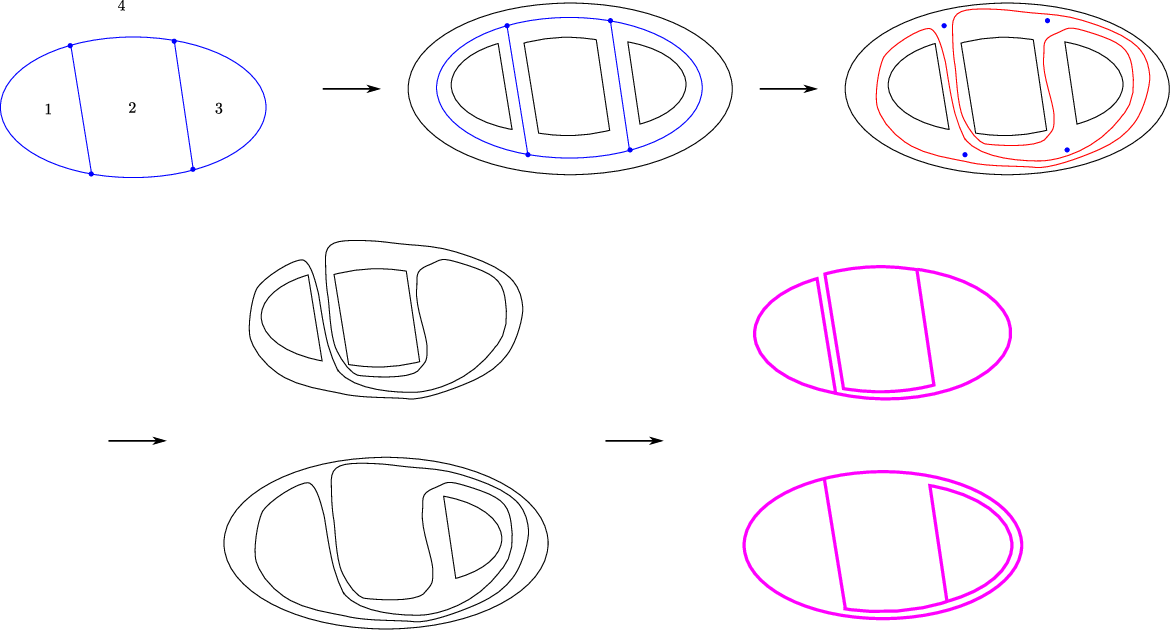}
\par\end{centering}
\caption{A pants decomposition of a metric ribbon graph. We get two dumbbell
graphs. \label{fig:A-pants-decomposition-1}}
\end{figure}

For the combinatorial moduli space, one can define pants decomposition
and Fenchel-Nielsen coordinates in a similar way. Let $G_{g,n,\mathbf{L}}\in\mathcal{M}_{g,n}^{\mathrm{comb}}(\mathbf{L})$
be a metric ribbon graph, whose boundary components are labeled by
indices $a=1,\cdots,n$ and have lengths $\mathbf{L}=(L_{1},\cdots,L_{n})$.
As mentioned above, one can decompose $G_{g,n,\mathbf{L}}$ into $2g-2+n$
combinatorial pairs of pants. This can be done without resorting to
the $\beta\to\infty$ limit of hyperbolic surfaces via the following
procedure. Roughly speaking, what we should do is to thicken $G_{g,n,\mathbf{L}}$,
cut it along $3g-3+n$ non boundary parallel (also called essential)
simple closed curves to get $2g-2+n$ pairs of pants, and shrink them
back to metric ribbon graphs (Figure \ref{fig:A-pants-decomposition}).
A simple closed curve on thickened $G_{g,n,\mathbf{L}}$ corresponds
to a non-backtracking closed curve $\gamma$ on $G_{g,n,\mathbf{L}}$,
which travels along its edges. In general, $\gamma$ visits some edges
of $G_{g,n,\mathbf{L}}$ multiple number of times (Figure \ref{fig:A-pants-decomposition-1}).
Given such a decomposition, it is easy to write down an equation like
(\ref{eq:identity}) for $\delta_{G_{g,n,\mathbf{L}}}$. 

\begin{figure}
\begin{centering}
\includegraphics[scale=0.75]{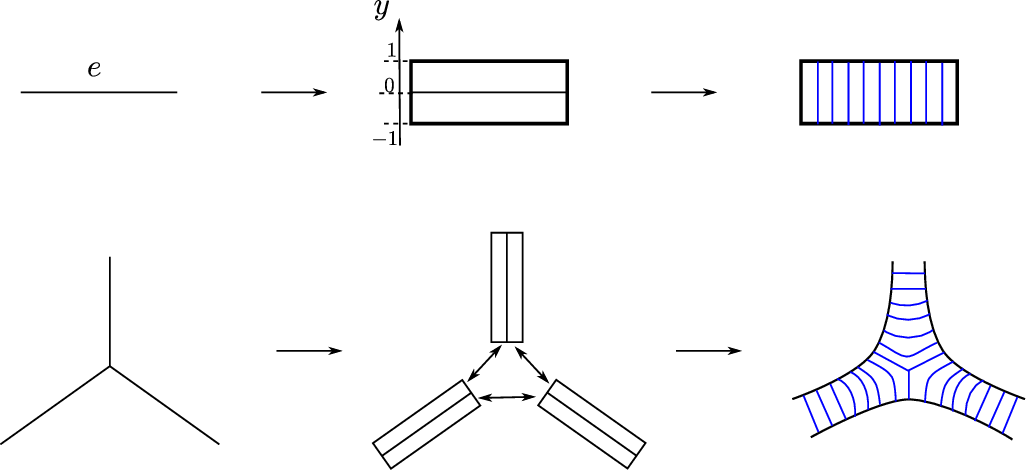}
\par\end{centering}
\caption{$\left|G_{g,n,\mathbf{L}}\right|$ and the measured foliation. \label{fig:-and-the}}

\end{figure}

The above intuitively explained procedure is defined precisely in
\cite{andersen2020kontsevich}. What we loosely call ``thickened
$G_{g,n,\mathbf{L}}$'' is given as the geometric realization $\left|G_{g,n,\mathbf{L}}\right|$
of $G_{g,n,\mathbf{L}}$ defined as follows. Given an edge $e$ of
$G_{g,n,\mathbf{L}}$, we thicken it by considering $e\times I$,
where $I=\left\{ y\,|\,-1\leq y\leq1\right\} $ is a closed interval.
We get a rectangle as depicted in Figure \ref{fig:-and-the} on which
we have coordinates $(x,y)$ where $x$ is a coordinate on $e$ satisfying
$dx=\sqrt{\varphi(z)}dz$ with $\varphi(z)dz^{2}$ being the Strebel
differential on $\mathrm{gr}_{\infty}(G_{g,n,\mathbf{L}})$. The rectangle
is foliated by leaves given by the curves $x=\text{constant}$. The
edge $e$ is identified with the curve $y=0$. At the vertices of
$G_{g,n,\mathbf{L}}$ where three edges meet, we glue
the thickened edges as illustrated in Figure \ref{fig:-and-the}.
The gluing rule is defined for other cases in a similar way. Gluing
all thickened edges, we get a genus $g$ surface with $n$ boundary
components which is denoted by $\left|G_{g,n,\mathbf{L}}\right|$.
The construction looks quite like that of Feynman diagrams in Witten's
open string field theory \cite{Witten1986}. We have a foliation with
isolated singularities on $\left|G_{g,n,\mathbf{L}}\right|$ 
equipped with the transverse measure $\left|dx\right|$, by which
we can define the lengths of arcs transverse to the foliation. Such
a foliation is called a measured foliation \cite{thurston1988geometry,fathi2021thurston,penner1992combinatorics}. 

The surface $\left|G_{g,n,\mathbf{L}}\right|$ can be cut along $3g-3+n$
essential simple closed curves so that we get $2g-2+n$ pairs of pants.
By deforming the curves homotopically and changing the foliation if
necessary\footnote{If $G_{g,n,\mathbf{L}}$ is not a trivalent graph, we may have to
perform Whitehead moves.}, we can take the curves to be transverse to the foliation. Then each
pair of pants inherits the measured foliation which is equivalent
to that of $\left|G_{0,3,\mathbf{L}}\right|$ for some $\mathbf{L}=(L_{1},L_{2},L_{3})$.
$L_{1},L_{2},L_{3}$ coincide with the combinatorial lengths of the
boundary components of the pair of pants, defined through the transverse
measure. If we shrink the pairs of pants down to the combinatorial
ones, we get a combinatorial pants decomposition of $G_{g,n,\mathbf{L}}$. 

Given a pants decomposition, the combinatorial Fenchel-Nielsen coordinates
$(l_{s}^{\mathrm{comb}};\tau_{s}^{\mathrm{comb}})\ (s=1,\cdots,3g-3+n)$
are defined by using the measured foliation on $\left|G_{g,n,\mathbf{L}}\right|$.
$l_{s}^{\mathrm{comb}}$ coincide with the combinatorial lengths of
the closed curves on $\left|G_{g,n,\mathbf{L}}\right|$, along which
$\left|G_{g,n,\mathbf{L}}\right|$ is cut. $\tau_{s}$ denote the
twist parameters which specify how boundaries of different pairs of
pants are identified on $\left|G_{g,n,\mathbf{L}}\right|$. Unlike
the Fenchel-Nielsen coordinates for hyperbolic surfaces, for fixed
$(l_{1}^{\mathrm{comb}},\cdots,l_{3g-3+n}^{\mathrm{comb}})\in\mathbb{R}_{+}^{3g-3+n}$,
some values of $(\tau_{1}^{\mathrm{comb}},\cdots,\tau_{3g-3+n}^{\mathrm{comb}})\in\mathbb{R}^{3g-3+n}$
are not admissible \cite{andersen2020kontsevich}. An example of such
a twist is presented in appendix \ref{sec:Non-admissible-twists}.
However, the non-admissible values of $\tau_{s}$ are of measure zero
and the coordinates $(l_{s}^{\mathrm{comb}};\tau_{s}^{\mathrm{comb}})$
with $0<l_{s}<\infty,-\infty<\tau_{s}<\infty$. are almost everywhere
global coordinates on the combinatorial version of the Teichm\"{u}ller
space $\mathcal{T}_{g,n}^{\mathrm{comb}}(\mathbf{L})$. The moduli
space $\mathcal{M}_{g,n}^{\mathrm{comb}}(\mathbf{L})$ is given as
\[
\mathcal{M}_{g,n}^{\mathrm{comb}}(\mathbf{L})=\mathcal{T}_{g,n}^{\mathrm{comb}}(\mathbf{L})/\mathrm{Mod}_{g,n}\,.
\]
Hence the combinatorial Fenchel-Nielsen coordinates $(l_{s}^{\mathrm{comb}};\tau_{s}^{\mathrm{comb}})$
can be used as local coordinates on $\mathcal{M}_{g,n}^{\mathrm{comb}}(\mathbf{L})$.

In \cite{Ishibashi2023}, a string field theory based on the pants
decomposition of hyperbolic surfaces was constructed. We would like
to construct a theory based on the pants decomposition of metric ribbon
graphs. Since the theory of $\mathcal{M}_{g,n}^{\mathrm{comb}}(\mathbf{L})$
can be developed in parallel to that of $\mathcal{M}_{g,n}(\mathbf{L})$
as we explained above, we will follow the procedure of \cite{Ishibashi2023}
closely, to attain our goal.

\section{Surface states \label{sec:The-surface-states}}

In order to construct a string field theory, let us first introduce surface states. 
%The expression (\ref{eq:deltaG}) is rather formal but one can make
%it tractable by introducing the surface states. 
As is described in
appendix \ref{sec:Combinatorial-moduli-space}, given $G_{g,n,\mathbf{L}}\in\mathcal{M}_{g,n}^{\mathrm{comb}}(\mathbf{L})$,
we can construct a punctured Riemann surface $\mathrm{gr}_{\infty}(G_{g,n,\mathbf{L}})$
by attaching semi-infinite cylinders $D_{a}$ to the boundary components
of $G_{g,n,\mathbf{L}}$. We take the local coordinate $u_{a}$ in (\ref{eq:uaz}) on
$D_{a}$.
In this way, from $G_{g,n,\mathbf{L}}$ we obtain a punctured Riemann
surface $\mathrm{gr}_{\infty}(G_{g,n,\mathbf{L}})$ with local coordinates
$u_{a}$ defined up to phases around the punctures $u_{a}=0$. Let
$\mathcal{H}$ denote the state space of the worldsheet theory. We
define the surface state $|G_{g,n,\mathbf{L}}\rangle\in\mathcal{H}^{\otimes n}$
associated to the punctured Riemann surface $\mathrm{gr}_{\infty}(G_{g,n,\mathbf{L}})$
with local coordinates $u_{a}$ to satisfy
\begin{equation}
_{1\cdots n}\langle G_{g,n,\mathbf{L}}|\Psi_{1}\rangle_{1}\cdots|\Psi_{n}\rangle_{n}=\langle\prod_{a=1}^{n}u_{a}^{-1}\circ\mathcal{O}_{\Psi_{a}}(0)\rangle_{\mathrm{gr}_{\infty}(G_{g,n,\mathbf{L}})}\, ,
\label{eq:surface}
\end{equation}
for any $|\Psi_{a}\rangle\in\mathcal{H}\  (a=1,\cdots ,n)$.
Here $\langle G_{g,n,\mathbf{L}}|$ denotes the BPZ conjugate of $|G_{g,n,\mathbf{L}}\rangle$,
$\mathcal{O}_{\Psi_{a}}$ denotes the operator corresponding to the
state $|\Psi_{a}\rangle$ and $\left\langle \cdot\right\rangle _{\Sigma}$
denotes the correlation function of the worldsheet theory on surface
$\Sigma$. Since $|G_{g,n,\mathbf{L}}\rangle$ is uniquely determined
by the values of $\langle G_{g,n,\mathbf{L}}|\Psi_{1}\rangle\cdots|\Psi_{n}\rangle$,
(\ref{eq:surface}) can be regarded as giving a definition of the
surface state $|G_{g,n,\mathbf{L}}\rangle$. 

The surface state $|G_{g,n,\mathbf{L}}\rangle\in\mathcal{H}^{\otimes n}$
is the state such that 
\begin{equation}
_{1\cdots n}\langle G_{g,n,\mathbf{L}}|X^{(1)}\rangle_{1}\cdots|X^{(n)}\rangle_{n}=\delta_{G_{g,n,\mathbf{L}}}\left[X^{(1)},\cdots,X^{(n)}\right]
\,,
\label{eq:surfacedelta}
\end{equation}
holds, where $|X\rangle$ denotes the coherent state of $X$. Indeed,
the right hand side of (\ref{eq:surface}) will be given
by 
\begin{equation}
\int\left[dX^{\prime (1)}\cdots dX^{\prime (n)}\right]\delta_{G_{g,n,\mathbf{L}}}\left[X^{\prime (1)},\cdots,X^{\prime (n)}\right]\Psi_{1}\left[X^{\prime (1)}\right]\cdots\Psi_{n}\left[X^{\prime (n)}\right]\,,
\end{equation}
where $\Psi_{a}\left[X^{\prime (a)}\right]=\langle X^{\prime (a)}|\Psi_{a}\rangle\ (a=1,\cdots,n)$.
Taking $|\Psi_{a}\rangle=|X^{(a)}\rangle $, we
obtain (\ref{eq:surfacedelta}). 

Using the surface state, (\ref{eq:identity}) can be recast in the
form
\begin{equation}
_{1234}\langle G|={}_{125}\langle G_{3}|{}_{634}\langle G_{3}^{\prime}|R_{56}\rangle\,,\label{eq:identity2}
\end{equation}
where $|R_{aa^{\prime}}\rangle\in\mathcal{H}^{\otimes2}$ denotes
the reflector satisfying
\[
_{a}\langle\phi|R_{aa^{\prime}}\rangle=|\phi\rangle_{a^{\prime}}\,,
\]
for any $|\phi\rangle\in\mathcal{H}$. For general $(g,n)$, we have
factorization identities of the forms
\begin{eqnarray}
_{1\cdots n}\langle G_{g,n,\mathbf{L}}| & = & _{1\cdots n_{1}\ -1}\langle G_{g_{1},n_{1}+1,\mathbf{L}_{1}}|\ {}_{0\ n_{1}+1\cdots n}\langle G_{g_{2},n_{2}+1,\mathbf{L}_{2}}|R_{-1\,0}\rangle\,,\nonumber \\
_{1\cdots n}\langle G_{g,n,\mathbf{L}}| & = & _{1\cdots n\ -1\,0}\langle G_{g-1,n+2,\mathbf{L^{\prime}}}|R_{-1\,0}\rangle\,,\label{eq:identity3}
\end{eqnarray}
with $g_{1}+g_{2}=g,\,n_{1}+n_{2}=n$. 

\subsection{Three string vertices}

In the string field theory we will construct, $\langle G_{g,n,\mathbf{L}}|$
is expressed in terms of three string vertices using the factorization
identities (\ref{eq:identity3}). The three string vertex corresponds
to a surface state $\langle G_{0,3,\mathbf{L}}|$ for $G_{0,3,\mathbf{L}}\in\mathcal{M}_{0,3}^{\mathrm{comb}}(\mathbf{L})$
with $\mathbf{L}=(L_{1},L_{2},L_{3})$. 
Since $\mathcal{M}_{0,3}^{\mathrm{comb}}(\mathbf{L})$ is a point, 
$\langle G_{0,3,\mathbf{L}}|$ is fixed by the values of $L_{1},L_{2},L_{3}$. 
$\langle G_{0,3,\mathbf{L}}|$
satisfies
\begin{equation}
\langle G_{0,3,\mathbf{L}}|\Psi_{1}\rangle|\Psi_{2}\rangle|\Psi_{3}\rangle=\langle\prod_{k=1}^{3}U_{k}^{-1}\circ\mathcal{O}_{\Psi_{k}}(0)\rangle_{\mathrm{gr}_{\infty}(G_{0,3,\mathbf{L}})}\,,\label{eq:G03L}
\end{equation}
where $U_{k}\ (k=1,2,3)$ are used for the $u_{a}$ in (\ref{eq:surface})
for later convenience. 

$\mathrm{gr}_{\infty}(G_{0,3,\mathbf{L}})$
can be expressed by the Riemann sphere $\hat{\mathbb{C}}$ with three
punctures. Let $z$ be the global complex coordinate on $\hat{\mathbb{C}}$.
We take the punctures to be at $z=z_{k}\ (k=1,2,3)$ with $(z_{1},z_{2},z_{3})=(0,1,\infty)$.
Let $\Phi(z)dz^{2}$ be the Strebel differential. It is fixed by the
conditions that
\[
\Phi(z)dz^{2}\sim\left(-\frac{1}{(2\pi)^{2}}\frac{L_{k}^{2}}{(z-z_{k})^{2}}+\cdots\right)dz^{2}\,,
\]
for $z\sim z_{k}$ and $\Phi(z)$ is holomorphic for $z\neq z_{k}$.
We get
\begin{eqnarray}
\Phi(z)dz^{2} & = & -\frac{1}{(2\pi)^{2}}\left[\frac{L_{1}^{2}}{z^{2}}+\frac{L_{2}^{2}}{(z-1)^{2}}+\frac{L_{3}^{2}-L_{1}^{2}-L_{2}^{2}}{z(z-1)}\right]dz^{2}\,,\nonumber \\
 & = & -\frac{L_{3}^{2}z^{2}+(L_{2}^{2}-L_{1}^{2}-L_{3}^{2})z+L_{1}^{2}}{(2\pi)^{2}z^{2}(z-1)^{2}}dz^{2}\,.\label{eq:Phi}
\end{eqnarray}
It is possible to show that 
\begin{equation}
\Phi(w)dw^{2}=\begin{cases}
\left.\Phi(z)dz^{2}\right|_{L_{1}\leftrightarrow L_{2}} & \text{ for }w=1-z\\
\left.\Phi(z)dz^{2}\right|_{L_{1}\leftrightarrow L_{3}} & \text{ for }w=\frac{1}{z}\\
\left.\Phi(z)dz^{2}\right|_{L_{2}\leftrightarrow L_{3}} & \text{ for }w=\frac{z}{z-1}
\end{cases}\,,\label{eq:Psiw}
\end{equation}
holds.

\begin{figure}
\begin{centering}
\includegraphics[scale=0.65]{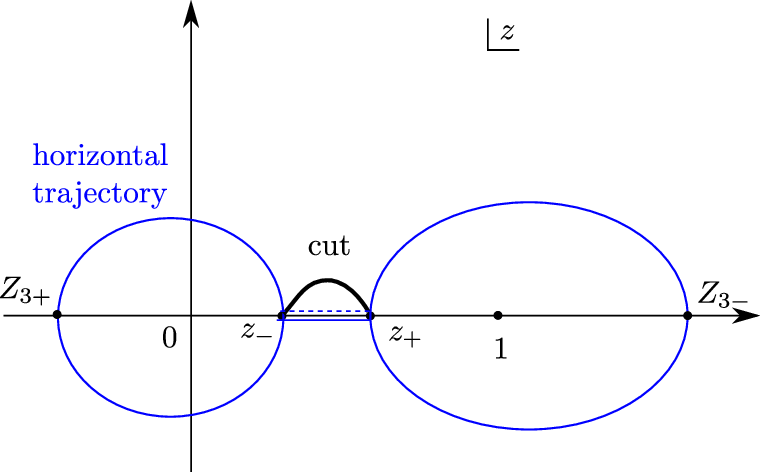}
\par\end{centering}
\caption{The case $D>0,\ 0<z_{-}<z_{+}<1$. \label{fig:The-case-.}}
\end{figure}

\begin{figure}
\begin{centering}
\includegraphics[scale=0.65]{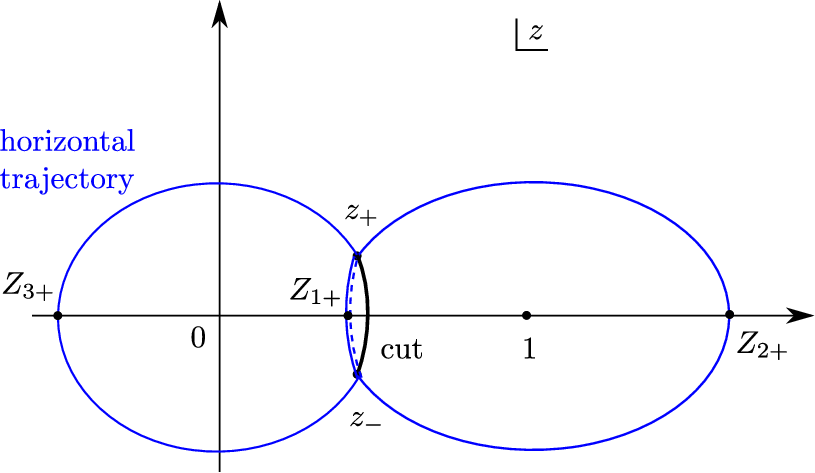}
\par\end{centering}
\caption{The case $D<0$. \label{fig:The-case-.-1}}
\end{figure}

The vertices of the metric ribbon graph $G_{0,3,\mathbf{L}}$ correspond
to the solutions of the equation 
\begin{equation}
\Phi(z)=0\,.\label{eq:interaction}
\end{equation}
There are generically two solutions $z=z_{\pm}$ to (\ref{eq:interaction})
where
\begin{equation}
z_{\pm}=\frac{L_{1}^{2}-L_{2}^{2}+L_{3}^{2}\pm\sqrt{D}}{2L_{3}^{2}}\,,\label{eq:zpm}
\end{equation}
with
\begin{equation}
D=-(L_{1}+L_{2}+L_{3})\prod_{k=1}^{3}(L_{1}+L_{2}+L_{3}-2L_{k})\,.\label{eq:D}
\end{equation}
$D<0,\,D=0$ and $D>0$ correspond to $\theta$ graph, figure-eight
graph and dumbbell graph depicted in Figure \ref{fig:The-combinatorial-pairs},
respectively. 

Since the state $\langle G_{0,3,\mathbf{L}}|$ is uniquely fixed by
the condition (\ref{eq:G03L}), we would like to calculate the local
coordinate $U_{k}$, which is equal to
\begin{equation}
e^{\pm\frac{2\pi i}{L_{k}}\int^{z}_{z_0}dz^{\prime}\sqrt{\Phi(z^{\prime})}}\,,\label{eq:ua}
\end{equation}
 up to a phase. The sign in the exponent should be chosen so that
$U_{k}=0$ corresponds to $z=z_{k}$. The explicit form of the integral
in the exponent is given by
\begin{eqnarray}
\int^{z}dz^{\prime}\sqrt{\Phi(z^{\prime})} & = & \frac{i}{2\pi}\left[-L_{1}\log\frac{(L_{2}^{2}-L_{1}^{2}-L_{3}^{2})z+2L_{1}^{2}-2L_{1}\sqrt{L_{3}^{2}z^{2}+(L_{2}^{2}-L_{1}^{2}-L_{3}^{2})z+L_{1}^{2}}}{z}\right.\nonumber \\
 &  & \quad\hphantom{\frac{1}{2\pi}}+L_{2}\log\frac{(L_{2}^{2}-L_{1}^{2}+L_{3}^{2})(z-1)+2L_{2}^{2}-2L_{2}\sqrt{L_{3}^{2}z^{2}+(L_{2}^{2}-L_{1}^{2}-L_{3}^{2})z+L_{1}^{2}}}{z-1}\nonumber \\
 &  & \quad\hphantom{\frac{1}{2\pi}}+L_{3}\log\left(2L_{3}^{2}z+L_{2}^{2}-L_{1}^{2}-L_{3}^{2}+2L_{3}\sqrt{L_{3}^{2}z^{2}+(L_{2}^{2}-L_{1}^{2}-L_{3}^{2})z+L_{1}^{2}}\right)\left.\vphantom{\frac{(L_{2}^{2}-L_{1}^{2}-L_{3}^{2})z+2L_{1}^{2}-2L_{1}\sqrt{L_{3}^{2}z^{2}+(L_{2}^{2}-L_{1}^{2}-L_{3}^{2})z+L_{1}^{2}}}{z}}\right]\,.\nonumber \\
 &  & \ \label{eq:introotphi}
\end{eqnarray}
We take the cut of the function 
\[
\sqrt{L_{3}^{2}z^{2}+(L_{2}^{2}-L_{1}^{2}-L_{3}^{2})z+L_{1}^{2}}\,,
\]
to be a curve which connects $z_{+}$ to $z_{-}$. 

Let us first consider the case $D>0$, which corresponds to either
$L_{1}+L_{2}<L_{3}$, $L_{1}+L_{3}<L_{2}$ or $L_{2}+L_{3}<L_{1}$.
For $L_{1}+L_{2}<L_{3}$ , $0<z_{-}<z_{+}<1$ holds and the cut can
be taken to be the one shown in Figure \ref{fig:The-case-.}. We obtain
\begin{equation}
\sqrt{L_{3}^{2}z^{2}+(L_{2}^{2}-L_{1}^{2}-L_{3}^{2})z+L_{1}^{2}}\sim\begin{cases}
-L_{1} & z\sim0\\
L_{2} & z\sim1\\
L_{3}z & z\sim\infty
\end{cases}\,,\label{eq:root}
\end{equation}
and
\begin{equation}
\int^{z}dz^{\prime}\sqrt{\Phi(z^{\prime})}\sim\begin{cases}
\frac{i}{2\pi}L_{1}\log z & z\sim0\\
\frac{i}{2\pi}L_{2}\log(z-1) & z\sim1\\
\frac{i}{2\pi}L_{3}\log z & z\sim\infty
\end{cases}\,.\label{eq:introotphising}
\end{equation}
Therefore we should take 
\begin{eqnarray}
U_{1} & = & e^{-\frac{2\pi i}{L_{1}}\int_{Z_{1}}^{z}dz^{\prime}\sqrt{\Phi(z^{\prime})}}\,,\nonumber \\
U_{2} & = & e^{-\frac{2\pi i}{L_{2}}\int_{Z_{2}}^{z}dz^{\prime}\sqrt{\Phi(z^{\prime})}}\,,\nonumber \\
U_{3} & = & e^{\frac{2\pi i}{L_{3}}\int_{Z_{3}}^{z}dz^{\prime}\sqrt{\Phi(z^{\prime})}}\,.\label{eq:Ua}
\end{eqnarray}
Here $Z_{1},Z_{2},Z_{3}$ are chosen to be $Z_{3+}$ or $z_{-}$,
$z_{+}$ or $Z_{3-}$, $Z_{3-}$ or $Z_{3+}$ in Figure \ref{fig:The-case-.}
respectively. Other cases can be dealt with in the same way or one
can use (\ref{eq:Psiw}) to get the formulas for $U_{k}$. 

Let us turn to the case $D<0$. Taking the cut shown in Figure \ref{fig:The-case-.-1},
we get (\ref{eq:Ua}) with $Z_{1},Z_{2},Z_{3}$ chosen to be $Z_{3+}$
or $Z_{1+}$, $Z_{1+}$ or $Z_{2+}$, $Z_{2+}$ or $Z_{3+}$ in the
figure respectively. $D=0$ is realized as a limiting case. 

\section{Off-shell amplitudes for closed strings\label{sec:The-off-shell-amplitudes}}
\subsection{$b$ ghost insertions\label{subsec:-ghost-insertions}}

Now let us define off-shell amplitudes in closed string theory 
%We would like to construct a theory in which the amplitudes are described
%by the contact interactions of strings represented by the critical
%graphs. 
utilizing the combinatorial moduli space
$\mathcal{M}_{g,n}^{\mathrm{comb}}(\mathbf{L})$ to describe the moduli
space of Riemann surfaces. The amplitudes will be expressed by integrals of the form
\[
\int_{\mathcal{M}_{g,n}^{\mathrm{comb}}(\mathbf{L})}\langle B_{6g-6+2n}\prod_{a=1}^{n}u_{a}^{-1}\circ\mathcal{O}_{\Psi_{a}}(0)\rangle_{\mathrm{gr}_{\infty}(G_{g,n,\mathbf{L}})}\,,
\]
where $\Psi_{a}$ are the external states. We take $B_{6g-6+2n}$ on the right
hand side to be
\begin{equation}
B_{6g-6+2n}=\prod_{s=1}^{3g-3+n}\left[b(\partial_{l_{s}^{\mathrm{comb}}})b(\partial_{\tau_{s}^{\mathrm{comb}}})\right]\bigwedge_{s=1}^{3g-3+n}\left[dl_{s}^{\mathrm{comb}}\wedge d\tau_{s}^{\mathrm{comb}}\right]\,.\label{eq:Bcomb}
\end{equation}
Here $(l_{s}^{\mathrm{comb}};\tau_{s}^{\mathrm{comb}})\ (s=1,\cdots,3g-3+n)$
are the combinatorial Fenchel-Nielsen coordinates on $\mathcal{M}_{g,n}^{\mathrm{comb}}(\mathbf{L})$
corresponding to a pants decomposition of $G_{g,n,\mathbf{L}}$. $b(\partial_{l_{s}^{\mathrm{comb}}})$
and $b(\partial_{\tau_{s}^{\mathrm{comb}}})$ are the $b$ ghost insertions
corresponding to the tangent vectors $\partial_{l_{s}^{\mathrm{comb}}}$
and $\partial_{\tau_{s}^{\mathrm{comb}}}$ on $\mathcal{M}_{g,n}^{\mathrm{comb}}(\mathbf{L})$
respectively.

In general, the $b$ ghost insertions are defined as follows \cite{Zwiebach1993,Sen2015,Lacroix2017,Erler2020,Erbin2021}.
Cutting the worldsheet along circles $\left\{ C_{A}\right\} $, we
decompose it into coordinate patches. We fix the orientation of $C_{A}$
so that $\sigma_{A}$ and $\sigma_{A}^{\prime}$ are the complex coordinates
on the left and the right of $C_{A}$ respectively. These two coordinates
are related by a transition function as
\[
\sigma_{A}=F_{A}(\sigma_{A}^{\prime})\,,
\]
where $F_{A}$ is a holomorphic function in a neighborhood of $C_{A}$.
Let $x_{t}\ (t=1,\cdots,6g-6+2n)$ be coordinates on the moduli space.
Then the $b$ ghost insertion corresponding to the tangent vector
$\partial_{x_{t}}$ is given by
\begin{equation}
b(\partial_{x_{t}})=\sum_{A}\left[\oint_{C_{\beta}}\frac{d\sigma_{A}}{2\pi i}\frac{\partial F_{A}}{\partial x_{t}}b(\sigma_{A})+\text{c.c.}\right]\,,\label{eq:general}
\end{equation}
where $\text{c.c.}$ denotes the antiholomorphic contribution. 

In our case, we decompose $\mathrm{gr}_{\infty}(G_{g,n,\mathbf{L}})$
into coordinate patches in the following way. Let us assume that $G_{g,n,\mathbf{L}}$
is a trivalent graph. For $\epsilon>0$, we define $\tilde{D}_{a}^{\epsilon}\subset D_{a}\ (a=1,\cdots,n)$
to be the region given by $\left|u_{a}\right|\leq e^{-\frac{2\pi}{L_{a}}\epsilon}$.
It is easy to see that $\mathrm{gr}_{\infty}(G_{g,n,\mathbf{L}})\backslash\cup_{a}\tilde{D}_{a}^{\epsilon}$
is homeomorphic to $\left|G_{g,n,\mathbf{L}}\right|$. Indeed $\mathrm{gr}_{\infty}(G_{g,n,\mathbf{L}})\backslash\cup_{a}\tilde{D}_{a}^{\epsilon}$
is made from rectangular neighborhoods of edges of $G_{g,n,\mathbf{L}}$
on which one can define coordinates $(x,y)\ (-1\leq y\leq1)$ such
that 
\begin{equation}
x+i\epsilon y=\int_{z_{0}}^{z}dz^{\prime}\sqrt{\varphi(z^{\prime})}\,,\label{eq:rect}
\end{equation}
where $z$ denotes a local coordinate on $\mathrm{gr}_{\infty}(G_{g,n,\mathbf{L}})\backslash\cup_{a}\tilde{D}_{a}^{\epsilon}$
and $z_{0}$ corresponds to a point on the edge. At the vertices the
rectangles are glued together as in Figure \ref{fig:-and-the}. With
the transverse measure $\left|dx\right|$, one can define a measured
foliation on $\mathrm{gr}_{\infty}(G_{g,n,\mathbf{L}})\backslash\cup_{a}\tilde{D}_{a}^{\epsilon}$
in the same way as in the case of $\left|G_{g,n,\mathbf{L}}\right|$.
Therefore we may identify $\mathrm{gr}_{\infty}(G_{g,n,\mathbf{L}})\backslash\cup_{a}\tilde{D}_{a}^{\epsilon}$
with $\left|G_{g,n,\mathbf{L}}\right|$ using the homeomorphism. 

\begin{figure}
\begin{centering}
\includegraphics[scale=0.75]{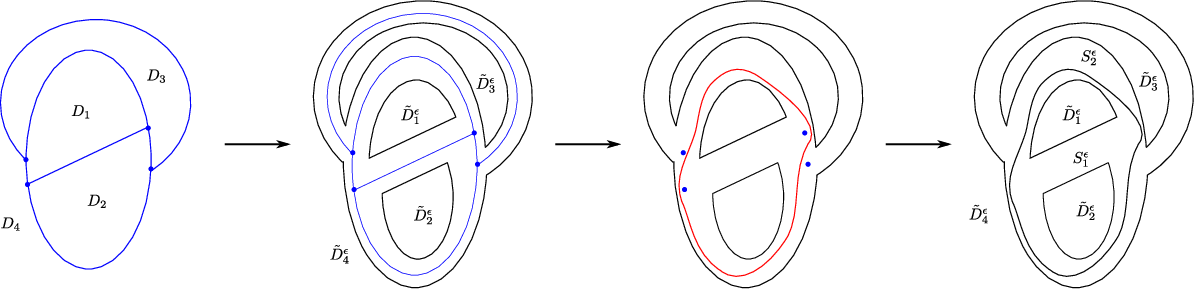}
\par\end{centering}
\caption{Coordinate patches on $\mathrm{gr}_{\infty}(G_{0,4,\mathbf{L}})$.\label{fig:Coordinate-patches-on}}

\end{figure}

The pants decomposition of $G_{g,n,\mathbf{L}}$ corresponding to
the Fenchel-Nielsen coordinates is obtained by cutting $\left|G_{g,n,\mathbf{L}}\right|$
along simple closed curves transverse to the foliation. By cutting
$\mathrm{gr}_{\infty}(G_{g,n,\mathbf{L}})\backslash\cup_{a}\tilde{D}_{a}^{\epsilon}$
along these curves, we can decompose it into $2g-2+n$ pairs of pants.
Let $S_{i}^{\epsilon}\ (i=1,\cdots,2g-2+n)$ denote 
these pairs of pants. $\mathrm{gr}_{\infty}(G_{g,n,\mathbf{L}})$
is decomposed into $S_{i}^{\epsilon}$ and $\tilde{D}_{a}^{\epsilon}$ (Figure \ref{fig:Coordinate-patches-on}).
The decomposition can be obtained by cutting $\mathrm{gr}_{\infty}(G_{g,n,\mathbf{L}})$
along $3g-3+2n$ circles. 

\begin{figure}
\begin{centering}
\includegraphics[scale=0.75]{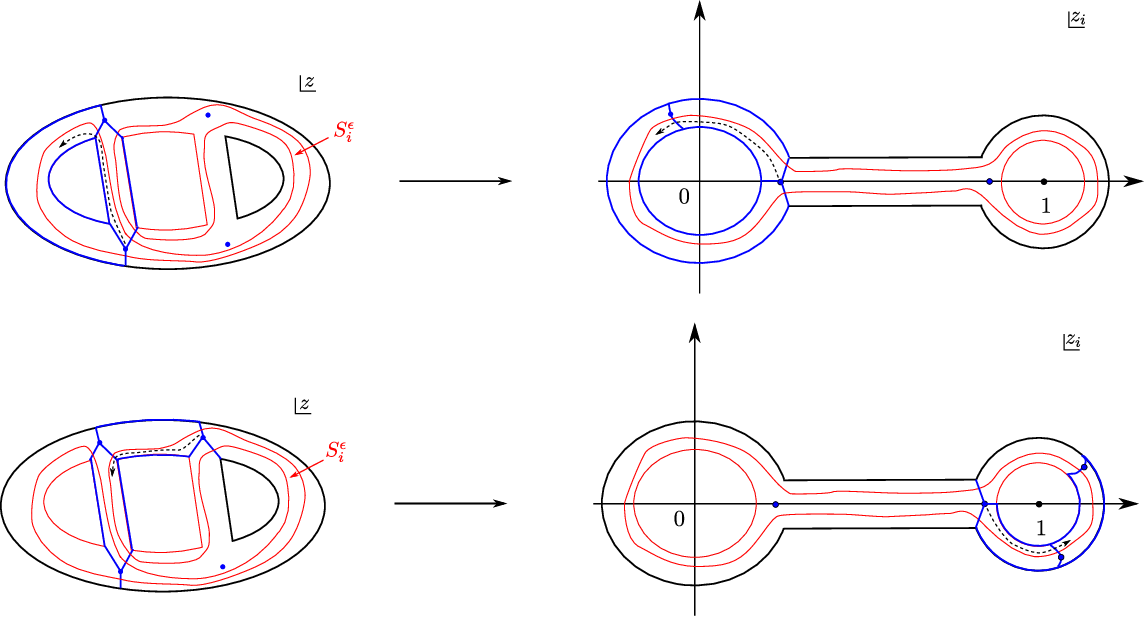}
\par\end{centering}
\caption{The analytic continuation of $z_{i}(z)$. The rectangular neighborhoods
are depicted as hexagons. \label{fig:The-map-.}}

\end{figure}

If the measured foliation on $S_{i}^{\epsilon}$ is equivalent to
that on $\left|G_{0,3,(L_{1},L_{2},L_{3})}\right|$, one can construct
a conformal map from $S_{i}^{\epsilon}$ to $\mathrm{gr}_{\infty}(G_{0,3,(L_{1},L_{2},L_{3})})$
in the following way. $\mathrm{gr}_{\infty}(G_{0,3,(L_{1},L_{2},L_{3})})$
can be expressed by the Riemann sphere $\hat{\mathbb{C}}$ with three
punctures. Let $z_{i}$ be the the global complex coordinate on the
Riemann sphere $\hat{\mathbb{C}}$ so that the Strebel differential
on $\mathrm{gr}_{\infty}(G_{0,3,(L_{1},L_{2},L_{3})})$ is given by
\[
\Phi(z_{i})dz_{i}^{2}\,,
\]
where $\Phi$ is defined in (\ref{eq:Phi}). Then we consider a conformal
map from $S_{i}^{\epsilon}$ to $\mathrm{gr}_{\infty}(G_{0,3,(L_{1},L_{2},L_{3})})$
such that the two zeros of $\varphi(z)dz^{2}$ in $S_{i}^{\epsilon}$
are mapped to the zeros\footnote{Which zeros are mapped to which is obvious from the topology of the
ribbon graphs. } of $\Phi(z_{i})dz_{i}^{2}$ and 
\begin{equation}
\sqrt{\varphi(z)}dz=\sqrt{\Phi(z_{i})}dz_{i}\,,\label{eq:rootphi}
\end{equation}
is satisfied. The map $z_{i}(z)$ is obtained by solving (\ref{eq:rootphi})
around the zeros of $\varphi(z)dz^{2}$ and analytically continuing
it to other regions in $S_{i}^{\epsilon}$. (\ref{eq:rootphi}) implies
that the rectangular neighborhoods of edges of $G_{g,n,\mathbf{L}}$
are mapped to those of edges of $G_{0,3,(L_{1},L_{2},L_{3})}$. As
can be seen from Figure \ref{fig:The-map-.}, the images of the neighborhoods
cover $\mathrm{gr}_{\infty}(G_{0,3,(L_{1},L_{2},L_{3})})\backslash\cup_{k=1}^{3}\tilde{D}_{k}^{\epsilon}$
and $S_{i}^{\epsilon}$ is conformally equivalent to a region in the
$z_{i}$ plane. 

Hence $z_{i}$ provides a local coordinate on $S_{i}^{\epsilon}$.
$u_{a}$ can be used as a local coordinate on $\tilde{D}_{a}^{\epsilon}$.
If $S_{i}^{\epsilon}\cap\tilde{D}_{a}^{\epsilon}=\partial\tilde{D}_{a}^{\epsilon}$,
$z_{i}$ and $u_{a}$ are related by
\begin{equation}
z_{i}=U_{k}^{-1}(u_{a})\,,\label{eq:fia}
\end{equation}
for some $k\ (k=1,2,3)$ in a neighborhood of $\partial\tilde{D}_{a}^{\epsilon}$.
In the same way, we can see that if $S_{i}^{\epsilon}\cap S_{j}^{\epsilon}=\mathrm{C}_{ij}^{\epsilon}$,
$z_{i}$ and $z_{j}$ are related by \footnote{This can be obtained by comparing $U_{k}(z_{i})$ and $U_{k^{\prime}}(z_{j})$
in a segment of $G_{g,n,\mathbf{L}}$ on which both of them are well-defined. }
\begin{equation}
z_{i}=U_{k}^{-1}\left(\frac{e^{i\theta_{ij}}}{U_{k^{\prime}}(z_{j})}\right)\,,\label{eq:Fij}
\end{equation}
for some $k,k^{\prime}$. Here $\theta_{ij}$ is the twist angle. 

Substituting (\ref{eq:fia}) and (\ref{eq:Fij}) into (\ref{eq:general}),
we obtain $b(\partial_{l_{s}^{\mathrm{comb}}})$ and $b(\partial_{\tau_{s}^{\mathrm{comb}}})$.
$l_{s}^{\mathrm{comb}}$ should be the combinatorial length of $\mathrm{C}_{ij}^{\epsilon}$
for some $i,j$. Hence $b(\partial_{l_{s}^{\mathrm{comb}}})$ can
be given by
\begin{eqnarray}
 &  & b(\partial_{l_{s}^{\mathrm{comb}}})=b_{S_{i}^{\epsilon}}(\partial_{l_{s}^{\mathrm{comb}}})+b_{S_{j}^{\epsilon}}(\partial_{l_{s}^{\mathrm{comb}}})\,,\nonumber \\
 &  & b_{S_{i}^{\epsilon}}(\partial_{l_{s}^{\mathrm{comb}}})=-\oint_{\partial S_{i}^{\epsilon}}\frac{dz_{i}}{2\pi i}\frac{\partial U_{k}(z_{i})}{\partial l_{s}^{\mathrm{comb}}}\left(\frac{\partial U_{k}(z_{i})}{\partial z_{i}}\right)^{-1}b(z_{i})+\text{c.c.}\,,\nonumber \\
 &  & b_{S_{j}^{\epsilon}}(\partial_{l_{s}^{\mathrm{comb}}})=-\oint_{\partial S_{j}^{\epsilon}}\frac{dz_{j}}{2\pi i}\frac{\partial U_{k}(z_{j})}{\partial l_{s}^{\mathrm{comb}}}\left(\frac{\partial U_{k}(z_{j})}{\partial z_{j}}\right)^{-1}b(z_{j})+\text{c.c.}\,.\label{eq:partialls}
\end{eqnarray}
Here $k\ (k=1,2,3)$ for $U_{k}$ in each term is chosen so that $U_{k}$
gives a good coordinate on the relevant component of the boundary.
With the choice of $Z_{k}$ in (\ref{eq:Ua}), $\partial_{\tau_{s}^{\mathrm{comb}}}=\frac{1}{l_{s}^{\mathrm{comb}}}\partial_{\theta_{ij}}$
and we have 
\begin{equation}
b(\partial_{\tau_{s}^{\mathrm{comb}}})=-\frac{2\pi}{l_{s}}\left[\oint_{\mathrm{C}_{ij}^{\epsilon}}\frac{dz_{i}}{2\pi i}iU_{k}(z_{i})\left(\frac{\partial U_{k}(z_{i})}{\partial z_{i}}\right)^{-1}b(z_{i})+\text{c.c.}\right]\,,\label{eq:partialtaus}
\end{equation}
where $k$ for $U_{k}$ is chosen so that $U_{k}(z_{i})$ is a good
coordinate on $\mathrm{C}_{ij}^{\epsilon}$. The contours run along
$\mathrm{C}_{ij}^{\epsilon}$ so that $S_{j}^{\epsilon}$ lies to
its left. 

In the same way, if $S_{i}^{\epsilon}\cap\tilde{D}_{a}^{\epsilon}=\partial\tilde{D}_{a}^{\epsilon}$,
we define 
\begin{equation}
b_{S_{i}^{\epsilon}}(\partial_{L_{a}})=-\oint_{\partial S_{i}^{\epsilon}}\frac{dz_{i}}{2\pi i}\frac{\partial U_{k}(z_{i})}{\partial L_{a}}\left(\frac{\partial U_{k}(z_{i})}{\partial z_{i}}\right)^{-1}b(z_{i})+\text{c.c.}\,,\label{eq:partialla}
\end{equation}
which play important roles in the following. All these formulas look
quite similar to those in the hyperbolic case \cite{Ishibashi2023}. 

Thus we have constructed $b$ ghost insertions assuming that $G_{g,n,\mathbf{L}}$
is a trivalent graph. When $G_{g,n,\mathbf{L}}$ is not trivalent,
it is not possible to define the coordinate patches $S_{i}^{\epsilon}$
and $\tilde{D}_{a}^{\epsilon}$. However, nontrivalent graphs can
be identified as limits of trivalent graphs as lengths of some of
the edges go to $0$. Therefore we define $b$ ghost insertions for
a nontrivalent graph by Eqs. (\ref{eq:partialls}), (\ref{eq:partialtaus})
and (\ref{eq:partialla}) in such a limit. %
\begin{comment}
If one expresses $b(\partial_{l_{s}^{\mathrm{comb}}})$ and $b(\partial_{\tau_{s}^{\mathrm{comb}}})$
using a local coordinate on $\mathrm{gr}_{\infty}(G_{g,n,\mathbf{L}})$,
they may look singular in the limit. 
\end{comment}
{} For trivalent graphs, these $b$ ghost insertions can be shown to
be equal to those using the Schiffer variation \cite{Zwiebach1993},
for example, by deforming the contours. Since those $b$ ghost insertions
are well-defined for nontrivalent graphs, they are well-defined in
the limit. 

For later use, we will show that (\ref{eq:partialls}), (\ref{eq:partialtaus})
and (\ref{eq:partialla}) can be expressed in terms of quantities
defined on the ribbon graph. We can do so by taking the $\epsilon\to0$
limit using the fact that the $b$ ghost insertions do not depend
on $\epsilon$. On the $z_{i}$ plane, for $l_{s}^{\mathrm{comb}}=L_{k^{\prime}}$,
$b(\partial_{\tau_{s}^{\mathrm{comb}}})$ is recast into
\[
\pm\oint_{\partial\tilde{D}_{k^{\prime}}^{\epsilon}}\frac{dz_{i}}{2\pi i}\frac{1}{\sqrt{\Phi(z_{i})}}b(z_{i})+\text{c.c.}\,,
\]
using (\ref{eq:Ua}), and it can be shown to be equal to
\begin{equation}
\lim_{\epsilon\to0}b(\partial_{\tau_{s}^{\mathrm{comb}}})=\pm\oint_{\partial D_{k^{\prime}}}\frac{dz_{i}}{2\pi i}\frac{1}{\sqrt{\Phi(z_{i})}}b(z_{i})+\text{c.c.}\,,\label{eq:tauepsilon0}
\end{equation}
by a contour deformation. The right hand side of (\ref{eq:tauepsilon0})
is given in terms of quantities defined on the ribbon graph. In the
same way, $b_{S_{i}^{\epsilon}}(\partial_{l_{s}^{\mathrm{comb}}})$
can be transformed into 
\begin{eqnarray*}
 &  & -\sum_{k=1}^{3}\oint_{\partial\tilde{D}_{k}^{\epsilon}}\frac{dz_{i}}{2\pi i}\frac{\partial}{\partial L_{k^{\prime}}}\left(\frac{1}{L_{k}}\int_{Z_{k}}^{z_{i}}dz^{\prime}\sqrt{\Phi(z^{\prime})}\right)\frac{L_{k}}{\sqrt{\Phi(z_{i})}}b(z_{i})+\text{c.c.}\\
 &  & \quad=-\sum_{k=1}^{3}\oint_{\partial\tilde{D}_{k}^{\epsilon}}\frac{dz_{i}}{2\pi i}\frac{\partial}{\partial L_{k^{\prime}}}\left(\int_{Z_{k}}^{z_{i}}dz^{\prime}\sqrt{\Phi(z^{\prime})}\right)\frac{1}{\sqrt{\Phi(z_{i})}}b(z_{i})\\
 &  & \hphantom{\quad=}+\frac{1}{L_{k^{\prime}}}\oint_{\partial\tilde{D}_{k^{\prime}}^{\epsilon}}\frac{dz_{i}}{2\pi i}\frac{1}{\sqrt{\Phi(z_{i})}}b(z_{i})\int_{Z_{k^{\prime}}}^{z_{i}}dz^{\prime}\sqrt{\Phi(z^{\prime})}\\
 &  & \hphantom{\quad=}+\text{c.c.}\,.
\end{eqnarray*}
In this case, the function $\frac{\partial}{\partial L_{k^{\prime}}}\left(\int_{Z_{k}}^{z_{i}}dz^{\prime}\sqrt{\Phi(z^{\prime})}\right)\frac{1}{\sqrt{\Phi(z_{i})}}$
in the integrand develops poles at $z_{i}=z_{\pm}$. This function
can be expressed as 
\begin{eqnarray*}
 &  & \frac{\partial}{\partial L_{k^{\prime}}}\left(\int_{Z_{k}}^{z_{i}}dz^{\prime}\sqrt{\Phi(z^{\prime})}\right)\frac{1}{\sqrt{\Phi(z_{i})}}\\
 &  & \quad=\frac{\partial z_{+}}{\partial L_{k^{\prime}}}\frac{a_{+}}{z_{i}-z_{+}}+\frac{\partial z_{-}}{\partial L_{k^{\prime}}}\frac{a_{-}}{z_{i}-z_{-}}+R_{k^{\prime}}(z_{i})\,,
\end{eqnarray*}
where $a_{\pm}$ are constants and $R_{k^{\prime}}(z_{i})$ is a function
which does not have poles at $z_{i}=z_{\pm}$. Using this formula,
we can show that $b_{S_{i}^{\epsilon}}(\partial_{l_{s}^{\mathrm{comb}}})$
is equal to 
\begin{eqnarray}
b_{S_{i}^{0}}(\partial_{l_{s}^{\mathrm{comb}}}) & \equiv & \lim_{\epsilon\to0}b_{S_{a}^{\epsilon}}(\partial_{l_{s}^{\mathrm{comb}}})\nonumber \\
 & = & a_{+}\frac{\partial z_{+}}{\partial L_{k^{\prime}}}b(z_{+})+a_{-}\frac{\partial z_{-}}{\partial L_{k^{\prime}}}b(z_{-})\nonumber \\
 &  & -\sum_{k=1}^{3}\oint_{\partial D_{k}}\frac{dz_{i}}{2\pi i}R_{k^{\prime}}(z_{i})b(z_{i})\nonumber \\
 &  & +\frac{1}{L_{k^{\prime}}}\oint_{\partial D_{k^{\prime}}}\frac{dz_{i}}{2\pi i}\frac{1}{\sqrt{\Phi(z_{i})}}b(z_{i})\int_{Z_{k^{\prime}}}^{z_{i}}dz^{\prime}\sqrt{\Phi(z^{\prime})}\nonumber \\
 &  & +\text{c.c.}\,.\label{eq:lepsilon0}
\end{eqnarray}
The right hand side of this equation consists of quantities defined
on the critical graph on the $z_{i}$ plane. Using $\eqref{eq:rootphi}$,
we can express the right hand sides of  (\ref{eq:tauepsilon0}) and
(\ref{eq:lepsilon0}) in terms of quantities defined on $G_{g,n,\mathbf{L}}$
as long as $G_{g,n,\mathbf{L}}$ is trivalent. $b_{S_{i}^{0}}(\partial_{L_{a}})$
is defined in the same way and can be shown to be equal to $b_{S_{i}^{\epsilon}}(\partial_{L_{a}})$. 

\subsection{The off-shell amplitudes}

Since $b(\partial_{l_{s}^{\mathrm{comb}}})$ and $b(\partial_{\tau_{s}^{\mathrm{comb}}})$
can be expressed in terms of quantities defined on $G_{g,n,\mathbf{L}}$,
one can represent $B_{6g-6+2n}$ in (\ref{eq:Bcomb}) as an operator
acting on the surface state $\langle G_{g,n,\mathbf{L}}|$. Let $\mathcal{H}_{0}$
denote the subspace of $\mathcal{H}$, which consists of the states
$|\Psi\rangle$ satisfying 
\begin{equation}
b_{0}^{-}|\Psi\rangle=(L_{0}-\bar{L}_{0})|\Psi\rangle=0\,,\label{eq:b0-}
\end{equation}
where $b_{0}^{\pm}\equiv b_{0}\pm\bar{b}_{0}$. In our setup, the
external lines for the off-shell amplitudes are labeled by a state
in $\mathcal{H}_{0}$ and the length of the string. We define $g$
loop $n$ string amplitudes for $g\geq0,\ n\geq1,\ 2g-2+n>0$ to be
\begin{equation}
A_{g,n}\left((|\Psi_{1}\rangle,L_{1}),\cdots,(|\Psi_{n}\rangle,L_{n})\right)=2^{-\delta_{g,1}\delta_{n,1}}(2\pi i)^{-3g+3-n}\int_{\mathcal{M}_{g,n}^{\mathrm{comb}}(\mathbf{L})}\langle G_{g,n,\mathbf{L}}|B_{6g-6+2n}|\Psi_{1}\rangle\cdots|\Psi_{n}\rangle\,,\label{eq:Agncomb}
\end{equation}
for $|\Psi_{a}\rangle\in\mathcal{H}_{0}\ (a=1,\cdots,n)$. The factor
$2^{-\delta_{g,1}\delta_{n,1}}$ is due to the $\mathbb{Z}_{2}$ symmetry
possessed by $G_{1,1,L}$. Since $\mathcal{M}_{g,n}^{\mathrm{comb}}(\mathbf{L})$
is homeomorphic to the moduli space $\mathcal{M}_{g,n}$ of punctured
Riemann surfaces $\mathrm{gr}_{\infty}(G_{g,n,\mathbf{L}})$, these
amplitudes coincide with the conventional amplitudes when $|\Psi_{a}\rangle$
are on-shell physical states. 

In the following, we will manipulate the expression (\ref{eq:Agncomb})
of the amplitudes and derive identities satisfied by them. Since the
moduli space $\mathcal{M}_{g,n}^{\mathrm{comb}}(\mathbf{L})$ is noncompact,
the amplitudes may suffer from divergences coming from the boundary.
What we would like to do is to construct a formalism using which we
obtain the expression on the right hand side of (\ref{eq:Agncomb}),
and discuss the behavior at the boundary later. In order to proceed,
we here assume that there exists a good regularization such that the
regularized integrands of the amplitudes go to zero rapidly at the
boundary of moduli space. In many cases, the amplitudes may be divergent
when the regularization is removed. 

\section{A recursion relation for the off-shell amplitudes\label{sec:A-recursion-relation}}

In calculating the amplitudes (\ref{eq:Agncomb}), $\mathcal{M}_{g,n}^{\mathrm{comb}}(\mathbf{L})$
should be realized as a fundamental domain of $\mathrm{Mod}_{g,n}$
in $\mathcal{T}_{g,n}^{\mathrm{comb}}(\mathbf{L})$ but there is no
concrete description of it for general $(g,n)$. In calculations of
the volumes $V_{g,n}(\mathbf{L})$ of the moduli spaces $\mathcal{M}_{g,n}(\mathbf{L})$
for hyperbolic surfaces, a similar problem was overcome by Mirzakhani's
integration scheme \cite{Mirzakhani2006,Mirzakhani2007}. Mirzakhani
used McShane identity \cite{McShane1991} and its generalization (Mirzakhani-McShane
identity) to unfold the integrals over $\mathcal{M}_{g,n}(\mathbf{L})$
and made the calculation possible. By doing so, she derived a recursion
relation satisfied by $V_{g,n}(\mathbf{L})$. 

There exists a generalization of Mirzakhani-McShane identity \cite{bennett2012topological,andersen2020kontsevich}
for combinatorial moduli space $\mathcal{M}_{g,n}^{\mathrm{comb}}(\mathbf{L})$,
which can be used to unfold integrals over $\mathcal{M}_{g,n}^{\mathrm{comb}}(\mathbf{L})$.
Applying this method to the integral on the right hand side of (\ref{eq:Agncomb}),
it is possible to derive a recursion relation satisfied by the off-shell
amplitudes. 

\subsection{Mirzakhani-McShane identity for combinatorial moduli space}

Given $G_{g,n,\mathbf{L}}\in\mathcal{M}_{g,n}^{\mathrm{comb}}(\mathbf{L})$,
let $\beta_{1},\cdots,\beta_{n}$ be the boundary components of $\left|G_{g,n,\mathbf{L}}\right|$
such that the combinatorial lengths of $\beta_{1},\cdots,\beta_{n}$
are $L_{1},\cdots,L_{n}$ respectively. The Mirzakhani-McShane identity
for combinatorial moduli space is given by \cite{andersen2020kontsevich,bennett2012topological}
\begin{equation}
1=\sum_{a=2}^{n}\sum_{\gamma\in\mathscr{C}_{a}}\mathsf{B}(L_{1},L_{a},l_{\gamma})+\frac{1}{2}\sum_{\left(\gamma,\delta\right)\in\mathscr{C}_{1}}\mathsf{C}(L_{1},l_{\gamma},l_{\delta})\,,\label{eq:MM}
\end{equation}
for $3g-3+n>0,\ (g,n)\neq(1,1)$ and 
\begin{equation}
1=\sum_{\gamma\in\mathscr{C}^{\mathrm{T}}}\mathsf{C}(L_{1},l_{\gamma},l_{\gamma})\,,\label{eq:MMT}
\end{equation}
for $(g,n)=(1,1)$. Here
\begin{eqnarray*}
\mathscr{C}_{a} & \equiv & \left\{ \begin{array}{l}
\text{the set of homotopy classes of essential simple closed curves }\gamma\text{ on }\text{\ensuremath{\left|G_{g,n,\mathbf{L}}\right|}}\\
\text{which bound a pair of pants together with the bondary components }\beta_{1}\text{ and }\beta_{a}
\end{array}\right\} \,,\\
\mathscr{C}_{1} & \equiv & \left\{ \begin{array}{l}
\text{the set of ordered pairs of homotopy classes of essential simple closed curves }\left(\gamma,\delta\right)\\
\text{on }\text{\ensuremath{\left|G_{g,n,\mathbf{L}}\right|}}\text{ which bound a pair of pants together with the bondary component }\beta_{1}
\end{array}\right\} \,,\\
\mathscr{C}^{\mathrm{T}} & \equiv & \left\{ \begin{array}{l}
\text{the set of homotopy classes of essential simple closed curves }\gamma\text{ on }\text{\ensuremath{\left|G_{1,1,L_{1}}\right|}}\end{array}\right\} \,,
\end{eqnarray*}
$l_{\gamma}$ denotes the combinatorial length of $\gamma$ and
\begin{eqnarray}
\mathsf{B}(L_{1},L_{2},L_{3}) & = & \frac{1}{2L_{1}}\left(\left[L_{1}-L_{2}-L_{3}\right]_{+}-\left[-L_{1}+L_{2}-L_{3}\right]_{+}+\left[L_{1}+L_{2}-L_{3}\right]_{+}\right)\nonumber \\
 & = & \begin{cases}
0 & L_{1}+L_{2}\leq L_{3}\\
\frac{1}{2L_{1}}\left(L_{1}+L_{2}-L_{3}\right) & \left|L_{1}-L_{2}\right|<L_{3}<L_{1}+L_{2}\\
1 & L_{3}\leq L_{2}-L_{1}\\
\frac{1}{L_{1}}(L_{1}-L_{3}) & L_{3}\leq L_{1}-L_{2}
\end{cases}\,,\nonumber \\
\mathsf{C}(L_{1},L_{2},L_{3}) & = & \frac{1}{L_{1}}\left[L_{1}-L_{2}-L_{3}\right]_{+}\nonumber \\
 & = & \begin{cases}
0 & L_{1}\leq L_{2}+L_{3}\\
\frac{1}{L_{1}}(L_{1}-L_{2}-L_{3}) & L_{2}+L_{3}<L_{1}
\end{cases}\,,\label{eq:BC}
\end{eqnarray}
with $\left[x\right]_{+}=\max\left\{ x,0\right\} $. Unlike the hyperbolic
case, only a finite number of terms on the right hand sides of (\ref{eq:MM})
and (\ref{eq:MMT}) are nonzero because of the definitions of $\mathsf{B}$
and $\mathsf{C}$. 

\subsection{A recursion relation for the off-shell amplitudes}

A recursion relation for the off-shell amplitudes can be derived by
following the same procedure as in \cite{Ishibashi2023}. Let us first
introduce a basis of the state space of the worldsheet theory. We
define $\mathcal{H}_{0}^{c}$ to be the subspace of $\mathcal{H}$
which consists of the states $|\Psi\rangle$ satisfying
\[
c_{0}^{-}|\Psi\rangle=(L_{0}-\bar{L}_{0})|\Psi\rangle=0\,,
\]
with $c_{0}^{\pm}=c_{0}\pm\bar{c}_{0}$. We choose the bases $\{|\varphi_{i}\rangle\}$
and $\{|\varphi_{i}^{c}\rangle\}$ of $\mathcal{H}_{0}$ and $\mathcal{H}_{0}^{c}$
respectively so that\footnote{The bases in this paper are different from those used in \cite{Ishibashi2023}. }
\begin{eqnarray}
\langle\varphi_{i}^{c}|\varphi_{j}\rangle & = & \delta_{ij}\,,\nonumber \\
\langle\varphi_{j}|\varphi_{i}^{c}\rangle & = & (-1)^{\left|\varphi_{i}\right|}\delta_{ij}\,,\nonumber \\
\sum_{i}|\varphi_{i}\rangle\langle\varphi_{i}^{c}| & = & \frac{1}{2}b_{0}^{-}c_{0}^{-}P\equiv P_{+}\,,\nonumber \\
\sum_{i}|\varphi_{i}^{c}\rangle\langle\varphi_{i}|(-1)^{\left|\varphi_{i}\right|} & = & \frac{1}{2}c_{0}^{-}b_{0}^{-}P\equiv P_{-}\,,\nonumber \\
P^{(1)}|R_{12}\rangle & = & P^{(2)}|R_{12}\rangle=\sum_{i}\left[|\varphi_{i}\rangle_{1}|\varphi_{i}^{c}\rangle_{2}+|\varphi_{i}^{c}\rangle_{1}|\varphi_{i}\rangle_{2}(-1)^{\left|\varphi_{i}\right|}\right]\,,\label{eq:PR}
\end{eqnarray}
hold. Here
\[
P=\int_{0}^{2\pi}\frac{d\theta}{2\pi}e^{i\theta(L_{0}-\bar{L}_{0})}\,.
\]

In order to derive a recursion relation, it is convenient to define
the following amplitudes:
\begin{equation}
A_{g,n}^{I_{1}\cdots I_{n}}\equiv2^{-\delta_{g,1}\delta_{n,1}}(2\pi i)^{-3g+3-n}
\int_{\mathcal{M}_{g,n}^{\mathrm{comb}}(\mathbf{L})}
\langle G_{g,n,\mathbf{L}}|B_{6g-6+2n}B_{\alpha_{1}}^{1}\cdots B_{\alpha_{n}}^{n}|\varphi_{i_{1}}^{\alpha_{1}}\rangle\cdots|\varphi_{i_{n}}^{\alpha_{n}}\rangle\,.\label{eq:Agn}
\end{equation}
Here the external lines are labeled by $I_{a}=(i_{a},\alpha_{a},L_{a})\ (a=1,\cdots,n)$.
The indices $\alpha_{a}$ take values $\pm$ and 
\[
|\varphi_{i_{a}}^{\alpha_{a}}\rangle=\begin{cases}
|\varphi_{i_{a}}\rangle & \alpha_{a}=+\\
|\varphi_{i_{a}}^{c}\rangle & \alpha_{a}=-
\end{cases}\,.
\]
$B_{\alpha_{a}}^{a}$is given by
\begin{equation}
B_{\alpha_{a}}^{a}\equiv\begin{cases}
1 & \alpha_{a}=+\\
b_{0}^{-(a)}b_{S_{a}^{0}}(\partial_{L_{a}}) & \alpha_{a}=-
\end{cases}\,.\label{eq:alpha}
\end{equation}
Here we define $S_{a}^{\epsilon}$ to be the pair of pants one of
whose boundary component coincides with $\partial\tilde{D}_{a}^{\epsilon}$
in a pants decomposition of $\mathrm{gr}_{\infty}(G_{g,n,\mathbf{L}})\backslash\cup_{a}\tilde{D}_{a}^{\epsilon}$.
$b_{S_{a}^{\epsilon}}(\partial_{L_{a}})$ depends on the choice of
the pants decomposition, because it corresponds to the variation $L_{a}\to L_{a}+\varepsilon$
with the Fenchel-Nielsen coordinates $l_{s}^{\mathrm{comb}},\tau_{s}^{\mathrm{comb}}$
fixed. However, $b_{S_{a}^{\epsilon}}(\partial_{L_{a}})B_{6g-6+2n}$
does not depend on the choice and the amplitudes in (\ref{eq:Agn})
are well-defined. 
(\ref{eq:Agncomb}) is the special case of (\ref{eq:Agn}) with $\alpha_a=+\ (a=1,\cdots ,n)$.

Now we would like to derive a recursion relation for $A_{g,n}^{I_{1}\cdots I_{n}}$.
In order to do so, let us multiply (\ref{eq:MM}) by
\[
(2\pi i)^{-3g+3-n}\langle G_{g,n,\mathbf{L}}|B_{6g-6+2n}B_{\alpha_{1}}^{1}\cdots B_{\alpha_{n}}^{n}|\varphi_{i_{1}}^{\alpha_{1}}\rangle\cdots|\varphi_{i_{n}}^{\alpha_{n}}\rangle\,,
\]
 and integrate over $\mathcal{M}_{g,n}^{\mathrm{comb}}(\mathbf{L})$.
We obtain
\begin{eqnarray}
 &  & (2\pi i)^{-3g+3-n}\int_{\mathcal{M}_{g,n}^{\mathrm{comb}}(\mathbf{L})}\langle G_{g,n,\mathbf{L}}|B_{6g-6+2n}B_{\alpha_{1}}^{1}\cdots B_{\alpha_{n}}^{n}|\varphi_{i_{1}}^{\alpha_{1}}\rangle\cdots|\varphi_{i_{n}}^{\alpha_{n}}\rangle\nonumber \\
 &  & \quad=\sum_{a=2}^{n}\sum_{\gamma\in\mathscr{C}_{a}}\int_{\mathcal{M}_{g,n}^{\mathrm{comb}}(\mathbf{L})}\mathsf{B}(L_{1},L_{a},l_{\gamma})(2\pi i)^{-3g+3-n}\langle G_{g,n,\mathbf{L}}|B_{6g-6+2n}B_{\alpha_{1}}^{1}\cdots B_{\alpha_{n}}^{n}|\varphi_{i_{1}}^{\alpha_{1}}\rangle\cdots|\varphi_{i_{n}}^{\alpha_{n}}\rangle\nonumber \\
 &  & \hphantom{\quad=}+\frac{1}{2}\sum_{\left(\gamma,\delta\right)\in\mathscr{C}_{1}}\int_{\mathcal{M}_{g,n}^{\mathrm{comb}}(\mathbf{L})}\mathsf{C}(L_{1},l_{\gamma},l_{\delta})(2\pi i)^{-3g+3-n}\langle G_{g,n,\mathbf{L}}|B_{6g-6+2n}B_{\alpha_{1}}^{1}\cdots B_{\alpha_{n}}^{n}|\varphi_{i_{1}}^{\alpha_{1}}\rangle\cdots|\varphi_{i_{n}}^{\alpha_{n}}\rangle\,,\nonumber \\
 &  & \ \label{eq:intMM}
\end{eqnarray}
for $(g,n)\neq(1,1)$. The left hand side is equal to $A_{g,n}^{I_{1}\cdots I_{n}}$.
We get what we want by rewriting the right hand side of (\ref{eq:intMM})
following the proof of Proposition 3.13 in \cite{andersen2020kontsevich}. 

Let us first study the term 
\begin{equation}
\sum_{\gamma\in\mathscr{C}_{a}}\int_{\mathcal{M}_{g,n}^{\mathrm{comb}}(\mathbf{L})}\mathsf{B}(L_{1},L_{a},l_{\gamma})(2\pi i)^{-3g+3-n}\langle G_{g,n,\mathbf{L}}|B_{6g-6+2n}B_{\alpha_{1}}^{1}\cdots B_{\alpha_{n}}^{n}|\varphi_{i_{1}}^{\alpha_{1}}\rangle\cdots|\varphi_{i_{n}}^{\alpha_{n}}\rangle\,,\label{eq:Ca}
\end{equation}
on the right hand side of (\ref{eq:intMM}). Using the fact that $\sum_{\gamma\in\mathscr{C}_{a}}\int_{\mathcal{M}_{g,n}^{\mathrm{comb}}(\mathbf{L})}$
can be regarded as an integration over the space
\[
\mathcal{M}_{g,n}^{\mathrm{comb},\gamma}(\mathbf{L})\equiv\left\{ \left.(G_{g,n,\mathbf{L}},\gamma)\,\right|\,G_{g,n,\mathbf{L}}\in\mathcal{M}_{g,n}^{\mathrm{comb}}(\mathbf{L}),\,\gamma\in\mathscr{C}_{a}\right\} \,,
\]
we obtain
\begin{eqnarray*}
 &  & \sum_{\gamma\in\mathscr{C}_{a}}\int_{\mathcal{M}_{g,n}^{\mathrm{comb}}(\mathbf{L})}\mathsf{B}(L_{1},L_{a},l_{\gamma})(2\pi i)^{-3g+3-n}\langle G_{g,n,\mathbf{L}}|B_{6g-6+2n}B_{\alpha_{1}}^{1}\cdots B_{\alpha_{n}}^{n}|\varphi_{i_{1}}^{\alpha_{1}}\rangle\cdots|\varphi_{i_{n}}^{\alpha_{n}}\rangle\\
 &  & \quad=\int_{\mathcal{M}_{g,n}^{\mathrm{comb},\gamma}(\mathbf{L})}\mathsf{B}(L_{1},L_{a},l_{\gamma})(2\pi i)^{-3g+3-n}\langle G_{g,n,\mathbf{L}}|B_{6g-6+2n}B_{\alpha_{1}}^{1}\cdots B_{\alpha_{n}}^{n}|\varphi_{i_{1}}^{\alpha_{1}}\rangle\cdots|\varphi_{i_{n}}^{\alpha_{n}}\rangle\,.
\end{eqnarray*}

For $(G_{g,n,\mathbf{L}},\gamma)\in\mathcal{M}_{g,n}^{\mathrm{comb},\gamma}(\mathbf{L})$,
by cutting $\left|G_{g,n,\mathbf{L}}\right|$ along a representative
of $\gamma$, we get a pair of pants and a surface which has a foliation
equivalent to that of $\left|G_{g,n-1,\mathbf{L^{\prime}}}^{\prime}\right|$
for $G_{g,n-1,\mathbf{L^{\prime}}}^{\prime}\in\mathcal{M}_{g,n-1}^{\mathrm{comb}}(\mathbf{L^{\prime}})$
with boundary components $\beta_{1}^{\prime},\cdots\beta_{n-1}^{\prime}$
such that
\[
\mathbf{L^{\prime}}=(l_{\beta_{1}^{\prime}},\cdots,l_{\beta_{n-1}^{\prime}})=(l_{\gamma},L_{2},\cdots,\hat{L}_{a},\cdots,L_{n})\,.
\]
 $\mathcal{M}_{g,n}^{\mathrm{comb},\gamma}(\mathbf{L})$ can be described
by the triple $(l_{\gamma},\tau_{\gamma},G_{g,n-1,\mathbf{L^{\prime}}}^{\prime})$
where $\tau_{\gamma}$ is the twist parameter corresponding to $\gamma$.
$\mathcal{M}_{g,n}^{\mathrm{comb},\gamma}(\mathbf{L})$ corresponds
to the region
\[
0<l_{\gamma}<\infty,\ 0\leq\tau_{\gamma}<l_{\gamma}\,,
\]
with non-admissible values of $\tau_{\gamma}$ excluded. We pick a
pants decomposition of $\left|G_{g,n,\mathbf{L}}\right|$ such that
one pair of pants has boundary components $\beta_{1},\beta_{a},\gamma$
and define the Fenchel-Nielsen coordinates $l_{s}^{\mathrm{comb}},\tau_{s}^{\mathrm{comb}}\ (s=1,\cdots,3g-3+n)$
so that $(l_{1}^{\mathrm{comb}},\tau_{1}^{\mathrm{comb}})=(l_{\gamma},\tau_{\gamma})$.
$l_{s}^{\mathrm{comb}},\tau_{s}^{\mathrm{comb}}\ (s=2,\cdots,3g-3+n)$
give local coordinates on $\mathcal{M}_{g,n-1}^{\mathrm{comb}}(\mathbf{L^{\prime}})$
and we get 
\begin{eqnarray}
 &  & \int_{\mathcal{M}_{g,n}^{\mathrm{comb},\gamma}(\mathbf{L})}\mathsf{B}(L_{1},L_{a},l_{\gamma})(2\pi i)^{-3g+3-n}\langle G_{g,n,\mathbf{L}}|B_{6g-6+2n}B_{\alpha_{1}}^{1}\cdots B_{\alpha_{n}}^{n}|\varphi_{i_{1}}^{\alpha_{1}}\rangle\cdots|\varphi_{i_{n}}^{\alpha_{n}}\rangle\nonumber \\
 &  & \quad=\int_{0}^{\infty}dl_{\gamma}\int_{0}^{l_{\gamma}}d\tau_{\gamma}\int_{\mathcal{M}_{g,n-1}^{\mathrm{comb}}(\mathbf{L^{\prime}})}\mathsf{B}(L_{1},L_{a},l_{\gamma})(2\pi i)^{-3g+3-n}\nonumber \\
 &  & \hphantom{\quad=\int_{0}^{\infty}dl_{\gamma}\int_{0}^{l_{\gamma}}d\tau_{\gamma}\int_{\mathcal{M}_{g,n-1}^{\mathrm{comb}}(\mathbf{L^{\prime}})}\quad}\times\langle G_{g,n,\mathbf{L}}|b(\partial_{l_{\gamma}})b(\partial_{\tau_{\gamma}})B_{6g-8+2n}^{\prime}B_{\alpha_{1}}^{1}\cdots B_{\alpha_{n}}^{n}|\varphi_{i_{1}}^{\alpha_{1}}\rangle\cdots|\varphi_{i_{n}}^{\alpha_{n}}\rangle\,,\label{eq:Mgamma}
\end{eqnarray}
with
\[
B_{6g-8+2n}^{\prime}=\prod_{s=2}^{3g-3+n}\left[b(\partial_{l_{s}^{\mathrm{comb}}})b(\partial_{\tau_{s}^{\mathrm{comb}}})\right]\bigwedge_{s=2}^{3g-3+n}\left[dl_{s}^{\mathrm{comb}}\wedge d\tau_{s}^{\mathrm{comb}}\right]\,.
\]
For $l_{\gamma}$ with 
\[
\mathsf{B}(L_{1},L_{a},l_{\gamma})\neq0\,,
\]
it is known that all the values of $\tau_{\gamma}\ (0\leq\tau_{\gamma}<l_{\gamma})$
are admissible (Corollary 2.33 in \cite{andersen2020kontsevich}). 

Since $G_{g,n,\mathbf{L}}$ can be decomposed into a combinatorial
pair of pants $G_{0,3,(L_{1},L_{a},l_{\gamma})}$ and $G_{g,n-1,\mathbf{L^{\prime}}}^{\prime}$,
the right hand side of (\ref{eq:Mgamma}) is transformed into 
\begin{eqnarray}
 &  & \varepsilon_{a}\int_{0}^{\infty}dl_{\gamma}\mathsf{B}(L_{1},L_{a},l_{\gamma})2^{\delta_{g,1}\delta_{n,2}}(2\pi i)^{-3g+4-n}\nonumber \\
 &  & \hphantom{\varepsilon_{a}\int_{0}^{\infty}dl_{\gamma}\mathsf{B}}\times\left[_{1a\ -1}\langle G_{0,3,(L_{1},L_{a},l_{\gamma})}|B_{\alpha_{1}}^{1}B_{\alpha_{a}}^{a}B_{-}^{-1}|\varphi_{i_{1}}^{\alpha_{1}}\rangle_{1}|\varphi_{i_{a}}^{\alpha_{a}}\rangle_{a}\vphantom{\int_{\mathcal{M}_{g,n-1}^{\mathrm{comb}}(\mathbf{L^{\prime}})}{}_{02\cdots\hat{a}\cdots n}\langle G_{g,n-1,\mathbf{L^{\prime}}}^{\prime}|B_{6g-8+2n}^{\prime}B_{-}^{0}\prod_{\substack{2\leq a^{\prime}\leq n\\
a^{\prime}\neq a
}
}B_{\alpha_{a^{\prime}}}^{a^{\prime}}|\varphi_{i_{2}}^{\alpha_{2}}\rangle_{2}\cdots\widehat{|\varphi_{i_{a}}^{\alpha_{a}}\rangle_{a}}\cdots|\varphi_{i_{n}}^{\alpha_{n}}\rangle_{n}}\right.\nonumber \\
 &  & \hphantom{\varepsilon_{a}\int_{0}^{\infty}dl_{\gamma}\mathsf{B}\times\qquad}\times\int_{\mathcal{M}_{g,n-1}^{\mathrm{comb}}(\mathbf{L^{\prime}})}{}_{02\cdots\hat{a}\cdots n}\langle G_{g,n-1,\mathbf{L^{\prime}}}^{\prime}|B_{6g-8+2n}^{\prime}\prod_{\substack{2\leq a^{\prime}\leq n\\
a^{\prime}\neq a
}
}B_{\alpha_{a^{\prime}}}^{a^{\prime}}|\varphi_{i_{2}}^{\alpha_{2}}\rangle_{2}\cdots\widehat{|\varphi_{i_{a}}^{\alpha_{a}}\rangle_{a}}\cdots|\varphi_{i_{n}}^{\alpha_{n}}\rangle_{n}\nonumber \\
 &  & \hphantom{\varepsilon_{a}\int_{0}^{\infty}dl_{\gamma}\mathsf{B}\times\quad}+{}_{1a\ -1}\langle G_{0,3,(L_{1},L_{a},l_{\gamma})}|B_{\alpha_{1}}^{1}B_{\alpha_{a}}^{a}|\varphi_{i_{1}}^{\alpha_{1}}\rangle_{1}|\varphi_{i_{a}}^{\alpha_{a}}\rangle_{a}\nonumber \\
 &  & \hphantom{\varepsilon_{a}\int_{0}^{\infty}dl_{\gamma}\mathsf{B}\times\qquad}\left.\times\int_{\mathcal{M}_{g,n-1}^{\mathrm{comb}}(\mathbf{L^{\prime}})}{}_{02\cdots\hat{a}\cdots n}\langle G_{g,n-1,\mathbf{L^{\prime}}}^{\prime}|B_{6g-8+2n}^{\prime}B_{-}^{0}\prod_{\substack{2\leq a^{\prime}\leq n\\
a^{\prime}\neq a
}
}B_{\alpha_{a^{\prime}}}^{a^{\prime}}|\varphi_{i_{2}}^{\alpha_{2}}\rangle_{2}\cdots\widehat{|\varphi_{i_{a}}^{\alpha_{a}}\rangle_{a}}\cdots|\varphi_{i_{n}}^{\alpha_{n}}\rangle_{n}\right]\nonumber \\
 &  & \hphantom{\varepsilon_{a}\int_{0}^{\infty}dl_{\gamma}\mathsf{B}}\times P^{(-1)}|R_{-1\,0}\rangle\,,\label{eq:Mgamma2}
\end{eqnarray}
using (\ref{eq:partialls}) and (\ref{eq:partialtaus}). Here
\[
\varepsilon_{a}=(-1)^{\left|\Psi_{a}\right|(\left|\Psi_{2}\right|+\cdots+\left|\Psi_{a-1}\right|)}\,,
\]
with $\left|\Psi\right|$ being the Grassmannality of $|\Psi\rangle$.%
\begin{comment}
$S^{\prime}$ denotes the coordinate patch on $\mathrm{gr}_{\infty}(G_{g,n-1,\mathbf{L^{\prime}}}^{\prime})$
which corresponds to the pair of pants which has $\beta_{1}^{\prime}$
as a boundary component.
\end{comment}

Substituting (\ref{eq:PR}) into (\ref{eq:Mgamma2}), (\ref{eq:Ca})
can be expressed in terms of the amplitudes (\ref{eq:Agn}). In order
to simplify the formula, we introduce the following notation. For
$X_{I}=X(i,\alpha,L)$ and $Y^{I}=Y(i,\alpha,L)$, we define
\[
X_{I}Y^{I}=X^{I}Y_{I}\equiv\sum_{i}\sum_{\alpha=\pm}\int_{0}^{\infty}dLX(i,\alpha,L)Y(i,\alpha,L)\,.
\]
We also define 
\begin{eqnarray*}
B^{I_{1}I_{2}I_{3}} & \equiv & \mathsf{B}(L_{1},L_{2},L_{3})\langle G_{0,3,(L_{1},L_{2},L_{3})}|B_{\alpha_{1}}^{1}B_{\alpha_{2}}^{2}B_{\alpha_{3}}^{3}|\varphi_{i_{1}}^{\alpha_{1}}\rangle_{1}|\varphi_{i_{2}}^{\alpha_{2}}\rangle_{2}|\varphi_{i_{3}}^{\alpha_{3}}\rangle_{3}\,,\\
C^{I_{1}I_{2}I_{3}} & \equiv & \mathsf{C}(L_{1},L_{2},L_{3})\langle G_{0,3,(L_{1},L_{2},L_{3})}|B_{\alpha_{1}}^{1}B_{\alpha_{2}}^{2}B_{\alpha_{3}}^{3}|\varphi_{i_{1}}^{\alpha_{1}}\rangle_{1}|\varphi_{i_{2}}^{\alpha_{2}}\rangle_{2}|\varphi_{i_{3}}^{\alpha_{3}}\rangle_{3}\,,\\
G_{I_{1}I_{2}} & \equiv & \delta(L_{1}-L_{2})\delta_{i_{1},i_{2}}\left[\delta_{\alpha_{1},+}\delta_{\alpha_{2},-}+\delta_{\alpha_{1},-}\delta_{\alpha_{2},+}(-1)^{\left|\varphi_{i_{1}}\right|}\right]\,.
\end{eqnarray*}
Then (\ref{eq:Mgamma2}) is equal to 
\begin{equation}
\varepsilon_{a}B^{I_{1}I_{a}I}G_{IJ}A_{g,n-1}^{JI_{2}\cdots\hat{I_{a}}\cdots I_{n}}\,.\label{eq:first}
\end{equation}

The term 
\begin{equation}
\sum_{\left(\gamma,\delta\right)\in\mathscr{C}_{1}}\int_{\mathcal{M}_{g,n}^{\mathrm{comb}}(\mathbf{L})}\mathsf{C}(L_{1},l_{\gamma},l_{\delta})(2\pi i)^{-3g+3-n}\langle G_{g,n,\mathbf{L}}|B_{6g-6+2n}B_{\alpha_{1}}^{1}\cdots B_{\alpha_{n}}^{n}|\varphi_{i_{1}}^{\alpha_{1}}\rangle\cdots|\varphi_{i_{n}}^{\alpha_{n}}\rangle\,,\label{eq:second}
\end{equation}
on the right hand side of (\ref{eq:intMM}) can be dealt with in the
same way. In this case, $\sum_{\left(\gamma,\delta\right)\in\mathscr{C}_{1}}\int_{\mathcal{M}_{g,n}^{\mathrm{comb}}(\mathbf{L})}$
can be regarded as an integration over the space
\[
\mathcal{M}_{g,n}^{\mathrm{comb},(\gamma,\delta)}(\mathbf{L})\equiv\left\{ \left.(G_{g,n,\mathbf{L}},\gamma,\delta)\,\right|\,G_{g,n,\mathbf{L}}\in\mathcal{M}_{g,n}^{\mathrm{comb}}(\mathbf{L}),\,(\gamma,\delta)\in\mathscr{C}_{1}\right\} \,.
\]
Cutting $\left|G_{g,n,\mathbf{L}}\right|$ along representatives of
$\gamma$ and $\delta$, we get a pair of pants along with a connected
surface or two connected surfaces. $\mathcal{M}_{g,n}^{\mathrm{comb},(\gamma,\delta)}(\mathbf{L})$
consists of components corresponding to these different configurations
of surfaces. Each component is described by a tuple of variables $(l_{\gamma},\tau_{\gamma},l_{\delta},\tau_{\delta},G_{g-1,n+1,\mathbf{L^{\prime}}}^{\prime})$
or $(l_{\gamma},\tau_{\gamma},l_{\delta},\tau_{\delta},G_{g_{1},n_{1},\mathbf{L}_{1}}^{\prime},G_{g_{2},n_{2},\mathbf{L}_{2}}^{\prime\prime})$
with 
\begin{eqnarray*}
 &  & 0<l_{\gamma}<\infty,\ 0\leq\tau_{\gamma}<l_{\gamma}\,,\\
 &  & 0<l_{\delta}<\infty,\ 0\leq\tau_{\delta}<l_{\delta}\,.
\end{eqnarray*}
It is known that for $l_{\gamma},l_{\delta}$ satisfying
\[
\mathsf{C}(L_{1},l_{\gamma},l_{\delta})\neq0\,,
\]
all the values of $\tau_{\gamma},\tau_{\delta}\ (0\leq\tau_{\gamma}<l_{\gamma},\ 0\leq\tau_{\delta}<l_{\delta})$
are admissible. We can express (\ref{eq:second}) in terms of the
amplitudes (\ref{eq:Agn}). 

Putting everything together, we can recast (\ref{eq:intMM}) into
the following form of recursion relation for $3g-3+n>0,\ (g,n)\neq(1,1)$:
\begin{eqnarray}
A_{g,n}^{I_{1}\cdots I_{n}} & = & \sum_{a=2}^{n}\varepsilon_{a}B^{I_{1}I_{a}J}G_{JI}A_{g,n-1}^{II_{2}\cdots\hat{I}_{a}\cdots I_{n}}\nonumber \\
 &  & +\frac{1}{2}C^{I_{1}J^{\prime}J}G_{JI}G_{J^{\prime}I^{\prime}}\left[A_{g-1,n+1}^{II^{\prime}I_{2}\cdots I_{n}}+\sum_{\text{stable}}\frac{\varepsilon_{\mathcal{I}_{1}\mathcal{I}_{2}}}{(n_{1}-1)!(n_{2}-1)!}A_{g_{1},n_{1}}^{I\mathcal{I}_{1}}A_{g_{2},n_{2}}^{I^{\prime}\mathcal{I}_{2}}\right]\,.\label{eq:recursion}
\end{eqnarray}
Here $\mathcal{I}_{1},\mathcal{I}_{2}$ denote ordered sets of indices
with $n_{1}-1,n_{2}-1$ elements respectively. The sum $\sum_{\text{stable}}$
means the sum over $g_{1},g_{2},n_{1},n_{2},\mathcal{I}_{1},\mathcal{I}_{2}$
such that
\begin{eqnarray}
g_{1}+g_{2} & = & g\,,\nonumber \\
n_{1}+n_{2} & = & n+1\,,\nonumber \\
\mathcal{I}_{1}\cup\mathcal{I}_{2} & = & \left\{ I_{2},\cdots,I_{n}\right\} \,,\nonumber \\
\mathcal{I}_{1}\cap\mathcal{I}_{2} & = & \phi\,,\nonumber \\
2g_{1}-2+n_{1} & > & 0\,,\nonumber \\
2g_{2}-2+n_{2} & > & 0\,.\label{eq:stable}
\end{eqnarray}
$\varepsilon_{\mathcal{I}_{1}\mathcal{I}_{2}}=\pm1$ is the sign which
will appear when we change the order of the product $II^{\prime}I_{2}\cdots I_{n}$
to $I\mathcal{I}_{1}I^{\prime}\mathcal{I}_{2}$, regarding the indices
as Grassmann numbers with Grassmannality of the corresponding string
state. For $(g,n)=(1,1)$, we can derive 
\begin{equation}
A_{1,1}^{I}=\frac{1}{2}C^{IJK}G_{JK}\,,\label{eq:recursion11}
\end{equation}
from (\ref{eq:MMT}), in the same way. 

With the recursion relation, the calculation of the amplitudes is
reduced to the base case
\begin{equation}
A_{0,3}^{I_{1}I_{2}I_{3}}=\langle G_{0,3,(L_{1},L_{2},L_{3})}|B_{\alpha_{1}}^{1}B_{\alpha_{2}}^{2}B_{\alpha_{3}}^{3}|\varphi_{i_{1}}^{\alpha_{1}}\rangle_{1}|\varphi_{i_{2}}^{\alpha_{2}}\rangle_{2}|\varphi_{i_{3}}^{\alpha_{3}}\rangle_{3}\,.\label{eq:A03}
\end{equation}
We can express $A_{g,n}^{I_{1}\cdots I_{n}}$ for any $g\geq0,\ n\geq1,\ 2g-2+n>0$
in terms of the three string vertices by solving (\ref{eq:recursion})
and (\ref{eq:recursion11}) with (\ref{eq:A03}). 

For later convenience, we introduce fictitious amplitudes
\[
A_{0,1}^{I}\equiv0,\ A_{0,2}^{I_{1}I_{2}}\equiv G^{I_{1}I_{2}}\,,
\]
and define the generating functional of the off-shell amplitudes
\begin{equation}
W[J]\equiv\sum_{g=0}^{\infty}\sum_{n=1}^{\infty}g_{\mathrm{s}}^{2g-2+n}\frac{1}{n!}J_{I_{n}}\cdots J_{I_{1}}A_{g,n}^{I_{1}\cdots I_{n}}\,.\label{eq:WA}
\end{equation}
Here
\[
G^{I_{1}I_{2}}=\delta(L_{1}-L_{2})\delta_{i_{1},i_{2}}\left[\delta_{\alpha_{1},+}\delta_{\alpha_{2},-}(-1)^{\left|\varphi_{i_{1}}\right|}+\delta_{\alpha_{1},-}\delta_{\alpha_{2},+}\right]\,,
\]
which satisfies
\[
G_{I_{1}I_{2}}G^{I_{2}I_{3}}=\delta_{I_{1}}^{\ I_{3}}=\delta_{i_{1},i_{3}}\delta_{\alpha_{1},\alpha_{3}}\delta(L_{1}-L_{3})\,.
\]
It is straightforward to show that the equations (\ref{eq:recursion}),
(\ref{eq:recursion11}) and (\ref{eq:A03}) are equivalent to the
following equation \cite{Ishibashi2023}:
\begin{eqnarray}
L\frac{\delta W[J]}{\delta J_{I}} & = & LJ_{I^{\prime}}G^{I^{\prime}I}\nonumber \\
 &  & +\frac{1}{2}g_{\mathrm{s}}V^{II^{\prime}I^{\prime\prime}}G_{I^{\prime\prime}K^{\prime\prime}}G_{I^{\prime}K^{\prime}}\left[\frac{\delta^{2}W[J]}{\delta J_{K^{\prime\prime}}\delta J_{K^{\prime}}}+\frac{\delta W[J]}{\delta J_{K^{\prime\prime}}}\frac{\delta W[J]}{\delta J_{K^{\prime}}}\right]\nonumber \\
 &  & +g_{\mathrm{s}}W^{II^{\prime}I^{\prime\prime}}G_{I^{\prime\prime}K^{\prime\prime}}J_{I^{\prime}}\frac{\delta W[J]}{\delta J_{K^{\prime\prime}}}(-1)^{\left|I\right|\left|I^{\prime}\right|}\,.\label{eq:WArecursion}
\end{eqnarray}
Here $\left|I\right|$ denotes the Grassmannality of $|\varphi_{i}\rangle$ and 
\begin{eqnarray*}
V^{I_{1}I_{2}I_{3}} & = & \begin{cases}
(L_{1}-L_{2}-L_{3})\langle G_{0,3,(L_{1},L_{2},L_{3})}|B_{\alpha_{1}}^{1}B_{\alpha_{2}}^{2}B_{\alpha_{3}}^{3}|\varphi_{i_{1}}^{\alpha_{1}}\rangle_{1}|\varphi_{i_{2}}^{\alpha_{2}}\rangle_{2}|\varphi_{i_{3}}^{\alpha_{3}}\rangle_{3} & L_{2}+L_{3}<L_{1}\\
0 & L_{1}<L_{2}+L_{3}
\end{cases}\,,\\
W^{I_{1}I_{2}I_{3}} & = & \begin{cases}
0 & L_{1}+L_{2}<L_{3}\\
(L_{1}+L_{2}-L_{3})\langle G_{0,3,(L_{1},L_{2},L_{3})}|B_{\alpha_{1}}^{1}B_{\alpha_{2}}^{2}B_{\alpha_{3}}^{3}|\varphi_{i_{1}}^{\alpha_{1}}\rangle_{1}|\varphi_{i_{2}}^{\alpha_{2}}\rangle_{2}|\varphi_{i_{3}}^{\alpha_{3}}\rangle_{3} & \left|L_{1}-L_{2}\right|<L_{3}<L_{1}+L_{2}\\
\min(L_{1},L_{2})\langle G_{0,3,(L_{1},L_{2},L_{3})}|B_{\alpha_{1}}^{1}B_{\alpha_{2}}^{2}B_{\alpha_{3}}^{3}|\varphi_{i_{1}}^{\alpha_{1}}\rangle_{1}|\varphi_{i_{2}}^{\alpha_{2}}\rangle_{2}|\varphi_{i_{3}}^{\alpha_{3}}\rangle_{3} & L_{3}<\left|L_{1}-L_{2}\right|
\end{cases}\,.
\end{eqnarray*}

\section{The Fokker-Planck formalism\label{sec:The-Fokker-Planck-formalism}}

Eq. (\ref{eq:WArecursion}) has the same form as Eq. (49) in \cite{Ishibashi2023}
for the hyperbolic case\footnote{The definition of $W_{A}[J]$ in eq.(48) in \cite{Ishibashi2023}
is not correct, The sum over $n$ should be from $1$ to $\infty$. }. Therefore it is quite easy to develop the Fokker-Planck formalism
\cite{Ishibashi:1993nq,Jevicki1994,Ishibashi:1993nqz,Ikehara1994,Ikehara1995}
for string fields using which we can compute the off-shell amplitudes
$A_{g,n}^{I_{1}\cdots I_{n}}$. 

\subsection{The Fokker-Planck formalism}

The Fokker-Planck formalism is described by introducing operators
$\hat{\phi}^{I},\hat{\pi}_{I}$ and states $|0\rangle\negthinspace\rangle,\langle\negthinspace\langle0|$
which satisfy 
\begin{eqnarray*}
[\hat{\pi}_{I},\hat{\phi}^{K}] & = & \delta_{I}^{\ K}\,,\\{}
[\hat{\pi}_{I},\hat{\pi}_{K}] & = & [\hat{\phi}^{I},\hat{\phi}^{K}]=0\,,\\
\langle\negthinspace\langle0|\hat{\phi}^{I} & = & \hat{\pi}_{I}|0\rangle\negthinspace\rangle=0\,,
\end{eqnarray*}
with
\[
[X^{I},Y^{K}]\equiv X^{I}Y^{K}-(-1)^{|I||K|}Y^{K}X^{I}\,,
\]
Using the Fokker-Planck Hamiltonian
\begin{eqnarray}
\hat{H} & = & -L\hat{\pi}_{I}\hat{\pi}_{I^{\prime}}G^{I^{\prime}I}+L\hat{\phi}^{I}\hat{\pi}_{I}\nonumber \\
 &  & -\frac{1}{2}g_{\mathrm{s}}V^{II^{\prime}I^{\prime\prime}}G_{I^{\prime\prime}K^{\prime\prime}}G_{I^{\prime}K^{\prime}}\hat{\phi}^{K^{\prime\prime}}\hat{\phi}^{K^{\prime}}\hat{\pi}_{I}\nonumber \\
 &  & -g_{\mathrm{s}}W^{II^{\prime}I^{\prime\prime}}G_{I^{\prime\prime}K^{\prime\prime}}\hat{\phi}^{K^{\prime\prime}}\hat{\pi}_{I^{\prime}}\hat{\pi}_{I}\,,\label{eq:FPH}
\end{eqnarray}
the correlation functions of $\phi^{I}$'s are defined to be
\begin{equation}
\langle\negthinspace\langle\phi^{I_{1}}\cdots\phi^{I_{n}}\rangle\negthinspace\rangle\equiv\lim_{\tau\to\infty}\langle\negthinspace\langle0|e^{-\tau\hat{H}}\hat{\phi}^{I_{1}}\cdots\hat{\phi}^{I_{n}}|0\rangle\negthinspace\rangle\,,\label{eq:corr}
\end{equation}
The correlation functions can be calculated perturbatively with respect
to $g_{\mathrm{s}}$. The connected correlation functions $\langle\negthinspace\langle\phi^{I_{1}}\cdots\phi^{I_{n}}\rangle\negthinspace\rangle^{\mathrm{c}}$
are expanded as
\[
\langle\negthinspace\langle\phi^{I_{1}}\cdots\phi^{I_{n}}\rangle\negthinspace\rangle^{\mathrm{c}}=\sum_{g=0}^{\infty}g_{\mathrm{s}}^{2g-2+n}\langle\negthinspace\langle\phi^{I_{1}}\cdots\phi^{I_{n}}\rangle\negthinspace\rangle_{g}^{\mathrm{c}}\,.
\]

It is possible to prove the equality
\begin{equation}
A_{g,n}^{I_{1}\cdots I_{n}}=\langle\negthinspace\langle\phi^{I_{1}}\cdots\phi^{I_{n}}\rangle\negthinspace\rangle_{g}^{\mathrm{c}}\,,\label{eq:corr=00003DA}
\end{equation}
exactly in the same way as in the hyperbolic case \cite{Ishibashi2023},
by showing that the Schwinger-Dyson equation for the correlation functions
(\ref{eq:corr}) is equivalent to (\ref{eq:WArecursion}). Thus the
Fokker-Planck Hamiltonian (\ref{eq:FPH}) made from kinetic terms
and three string interaction terms, describes the string theory. 

A few comments are in order:
\begin{itemize}
\item (\ref{eq:WArecursion}) implies that
\begin{equation}
\left[\lim_{\tau\to\infty}\langle\negthinspace\langle0|e^{-\tau\hat{H}}\right]\hat{\mathcal{T}}^{I}=0\,,\label{eq:SD}
\end{equation}
holds with
\begin{eqnarray*}
\hat{\mathcal{T}}^{I} & \equiv & -L\hat{\pi}_{I^{\prime}}G^{II^{\prime}}+L\hat{\phi}^{I}\\
 &  & -\frac{1}{2}g_{\mathrm{s}}V^{II^{\prime}I^{\prime\prime}}G_{I^{\prime\prime}K^{\prime\prime}}G_{I^{\prime}K^{\prime}}\hat{\phi}^{K^{\prime\prime}}\hat{\phi}^{K^{\prime}}\\
 &  & -g_{\mathrm{s}}W^{II^{\prime}I^{\prime\prime}}G_{I^{\prime\prime}K^{\prime\prime}}\hat{\phi}^{K^{\prime\prime}}\hat{\pi}_{I^{\prime}}\,.
\end{eqnarray*}
The Fokker-Planck Hamiltonian can be written as
\[
\hat{H}=\hat{\mathcal{T}}^{I}\hat{\pi}_{I}\,.
\]
\item One may try to describe the theory using a path integral with action
$S[\phi^{I}]$ such that
\[
\langle\negthinspace\langle\phi^{I_{1}}\cdots\phi^{I_{n}}\rangle\negthinspace\rangle=\frac{\int[d\phi^{I}]e^{-S[\phi^{I}]}\phi^{I_{1}}\cdots\phi^{I_{n}}}{\int[d\phi^{I}]e^{-S[\phi^{I}]}}\,.
\]
In the same way as in the hyperbolic case, one can derive an equation
for $S[\phi^{I}]$ :
\begin{eqnarray}
 &  & [LG^{IJ}+g_{\mathrm{s}}W^{IJI^{\prime}}G_{I^{\prime}J^{\prime}}\phi^{J^{\prime}}]\frac{\delta S}{\delta\phi^{J}}\nonumber \\
 &  & \quad=L\phi^{I}-\frac{1}{2}g_{\mathrm{s}}V^{II^{\prime}I^{\prime\prime}}G_{I^{\prime}J^{\prime}}G_{I^{\prime\prime}J^{\prime\prime}}\phi^{J^{\prime\prime}}\phi^{J^{\prime}}+g_{\mathrm{s}}W^{IJI^{\prime}}G_{I^{\prime}I^{\prime\prime}}\,.\label{eq:Seq}
\end{eqnarray}
As in the hyperbolic case, this equation is not well-defined because
the last term on the right hand side is divergent because the integration
region includes infinitely many fundamental domains of the mapping
class group. Formally, it is possible to solve (\ref{eq:Seq}) perturbatively
and obtain $S[\phi^{I}]$ which will be nonpolynomial and divergent. 
\end{itemize}

\subsection{SFT notation}

It is convenient to rewrite the Fokker-Planck Hamiltonian in terms
of variables $|\phi^{\alpha}(L)\rangle,|\pi_{\alpha}(L)\rangle$ defined
by 
\begin{eqnarray*}
|\phi^{\alpha}(L)\rangle & = & \sum_{i}\hat{\phi}^{I}|\varphi_{i}^{-\alpha}\rangle(\delta_{\alpha,+}+\delta_{\alpha,-}(-1)^{\left|\varphi_{i}\right|})\,,\\
|\pi_{\alpha}(L)\rangle & = & \sum_{i}|\varphi_{i}^{\alpha}\rangle\hat{\pi}_{I}\,.
\end{eqnarray*}
 These fields take values in the Hilbert space of strings as is usually
the case in string field theory. They are Grassmann even and satisfy
the canonical commutation relations
\begin{eqnarray*}
[|\pi_{\alpha}(L)\rangle_{1},|\phi^{\alpha^{\prime}}(L^{\prime})\rangle_{2}] & = & \delta_{\alpha}^{\alpha^{\prime}}\delta(L-L^{\prime})P_{\alpha}^{(1)}|R_{12}\rangle\,,\\{}
[|\pi_{\alpha}(L)\rangle_{1},|\pi_{\alpha^{\prime}}(L^{\prime})\rangle_{2}] & = & [|\phi^{\alpha}(L)\rangle_{1},|\phi^{\alpha^{\prime}}(L^{\prime})\rangle_{2}]=0\,,
\end{eqnarray*}
and
\[
\langle\negthinspace\langle0||\phi^{\alpha}(L)\rangle=|\pi_{\alpha}(L)\rangle|0\rangle\negthinspace\rangle=0\,.
\]

In terms of $|\phi^{\alpha}(L)\rangle,|\pi_{\alpha}(L)\rangle$, the
Fokker-Planck Hamiltonian is expressed as
\begin{eqnarray*}
\hat{H} & = & \int_{0}^{\infty}dLL\left[\langle R_{12}|\phi^{\alpha}(L)\rangle_{1}|\pi_{\alpha}(L)\rangle_{2}-\langle R_{12}|\pi_{\alpha}(L)\rangle_{1}|\pi_{-\alpha}(L)\rangle_{2}\right]\\
 &  & -\frac{1}{2}g_{\mathrm{s}}\int_{0}^{\infty}dL_{3}\int_{0}^{L_{3}}dL_{1}\int_{0}^{L_{3}-L_{1}}dL_{2}(L_{3}-L_{1}-L_{2})\langle G_{0,3,\mathbf{L}}|B_{-\alpha_{1}}^{1}B_{-\alpha_{2}}^{2}B_{\alpha_{3}}^{3}|\phi^{\alpha_{1}}(L_{1})\rangle_{1}|\phi^{\alpha_{2}}(L_{2})\rangle_{2}|\pi_{\alpha_{3}}(L_{3})\rangle_{3}\\
 &  & -g_{\mathrm{s}}\int_{0}^{\infty}dL_{2}\int_{0}^{\infty}dL_{3}\int_{\left|L_{2}-L_{3}\right|}^{L_{2}+L_{3}}dL_{1}(L_{2}+L_{3}-L_{1})\langle G_{0,3,\mathbf{L}}|B_{-\alpha_{1}}^{1}B_{\alpha_{2}}^{2}B_{\alpha_{3}}^{3}|\phi^{\alpha_{1}}(L_{1})\rangle_{1}|\pi_{\alpha_{2}}(L_{2})\rangle_{2}|\pi_{\alpha_{3}}(L_{3})\rangle_{3}\\
 &  & -g_{\mathrm{s}}\int_{0}^{\infty}dL_{2}\int_{0}^{\infty}dL_{3}\int_{0}^{\left|L_{2}-L_{3}\right|}dL_{1}\min(L_{2},L_{3})\langle G_{0,3,\mathbf{L}}|B_{-\alpha_{1}}^{1}B_{\alpha_{2}}^{2}B_{\alpha_{3}}^{3}|\phi^{\alpha_{1}}(L_{1})\rangle_{1}|\pi_{\alpha_{2}}(L_{2})\rangle_{2}|\pi_{\alpha_{3}}(L_{3})\rangle_{3}\,,
\end{eqnarray*}
where $\mathbf{L}=(L_{1},L_{2},L_{3})$ and the sum over repeated
indices $\alpha_{1},\alpha_{2},\alpha_{3}$ is understood. $\hat{\phi}^{I}$
and $\hat{\pi}_{I}$ are given by
\begin{eqnarray*}
\hat{\phi}^{I} & = & \langle\varphi_{i}^{\alpha}|\phi^{\alpha}(L)\rangle\,,\\
\hat{\pi}_{I} & = & \langle\varphi_{i}^{-\alpha}|\pi_{\alpha}(L)\rangle(\delta_{\alpha,+}+\delta_{\alpha,-}(-1)^{\left|\varphi_{i}\right|})\,,
\end{eqnarray*}
and the connected correlation functions are written as
\begin{eqnarray*}
 &  & \langle\negthinspace\langle|\phi^{\alpha_{1}}(L_{1})\rangle_{1}\cdots|\phi^{\alpha_{n}}(L_{n})\rangle_{n}\rangle\negthinspace\rangle_{g}^{\mathrm{c}}\\
 &  & \quad=2^{-\delta_{g,1}\delta_{n,1}}(2\pi i)^{-3g+3-n}
 \int_{\mathcal{M}_{g,n}^{\mathrm{comb}}(\mathbf{L})}
 {}_{1^{\prime}\cdots n^{\prime}}\langle G_{g,n,\mathbf{L}}|B_{6g-6+2n}B_{\alpha_{1}}^{1^{\prime}}\cdots B_{\alpha_{n}}^{n^{\prime}}P_{-\alpha_{1}}^{(1)}|R_{1^{\prime}1}\rangle\cdots P_{-\alpha_{n}}^{(n)}|R_{n^{\prime}n}\rangle\,.
\end{eqnarray*}

The correlation functions can be expressed in the path integral formalism
as 
\begin{equation}
\langle\negthinspace\langle|\phi^{\alpha_{1}}(L_{1})\rangle\cdots|\phi^{\alpha_{n}}(L_{n})\rangle\rangle\negthinspace\rangle=\frac{\int[d\pi d\phi]e^{-I}|\phi^{\alpha_{1}}(0,L_{1})\rangle\cdots|\phi^{\alpha_{n}}(0,L_{n})\rangle}{\int[d\pi d\phi]e^{-I}}\,,\label{eq:Icorr}
\end{equation}
using the Euclidean action
\begin{equation}
I=\int_{0}^{\infty}d\tau\left[-\int_{0}^{\infty}dL\langle R|\pi_{\alpha}(\tau,L)\rangle\frac{\partial}{\partial\tau}|\phi^{\alpha}(\tau,L)\rangle+H(\tau)\right]\,.\label{eq:Eucaction}
\end{equation}
In (\ref{eq:Icorr}), $|\phi^{\alpha}(\tau,L)\rangle,|\pi_{\alpha}(\tau,L)\rangle$
are taken to satisfy the boundary conditions 
\[
\lim_{\tau\to\infty}|\phi^{\alpha}(\tau,L)\rangle=|\pi_{\alpha}(0,L)\rangle=0\,,
\]
and the reality condition \cite{Zwiebach1993,Sen2016}
\begin{eqnarray}
|\phi^{+}(\tau,L)\rangle^{\dagger} & = & \langle\phi^{+}(\tau,L)|\,,\nonumber \\
|\phi^{-}(\tau,L)\rangle^{\dagger} & = & -\langle\phi^{-}(\tau,L)|\,.\label{eq:reality}
\end{eqnarray}

\subsection{BRST invariant formulation}

As in the hyperbolic case, the amplitudes defined in (\ref{eq:Agn})
are invariant under the BRST transformation
\begin{eqnarray*}
\delta_{\eta}|\phi^{+}(L)\rangle & = & \eta P_{-}Q|\phi^{+}(L)\rangle\,,\\
\delta_{\eta}|\phi^{-}(L)\rangle & = & \eta Q|\phi^{-}(L)\rangle-\eta b_{0}^{-}P\partial_{L}|\phi^{+}(L)\rangle\,,
\end{eqnarray*}
where $Q$ is the BRST operator of the worldsheet theory and $\eta$
is a Grassmann odd parameter. In the Fokker-Planck formalism the generator
of the transformation is given by
\begin{equation*}
\hat{Q}  =  \int dL\left[\langle R_{12}|Q|\phi^{\alpha}(L)\rangle_{1}|\pi_{\alpha}(L)\rangle_{2}
-\langle R_{12}|b_{0}^{-}P\partial_{L}|\phi^{+}(L)\rangle_{1}|\pi_{-}(L)\rangle_{2}\right]\,.
\end{equation*}

Although $\hat{Q}$ satisfies
\begin{eqnarray*}
 &  & \hat{Q}|0\rangle\negthinspace\rangle=\langle\negthinspace\langle0|\hat{Q}=0\,,\\
 &  & \left[\lim_{\tau\to\infty}\langle\negthinspace\langle0|e^{-\tau\hat{H}}\right]\hat{Q}=0\,,
\end{eqnarray*}
the Fokker-Planck Hamiltonian $\hat{H}$ is not invariant under the
BRST transformation. Since $\hat{H}$ can be written as 
\[
\hat{H}=\int_{0}^{\infty}dL\langle R_{12}|\mathcal{T}^{\alpha}(L)\rangle_{1}|\pi_{\alpha}(L)\rangle_{2}\,,
\]
with 
\[
|\mathcal{T}^{\alpha}(L)\rangle=\sum_{i}\hat{T}^{I}|\varphi_{i}^{-\alpha}\rangle(\delta_{\alpha,+}+\delta_{\alpha,-}(-1)^{\left|\varphi_{i}\right|})\,,
\]
the BRST variation of $\hat{H}$ is given by
\[
[\hat{Q},\hat{H}]=\int_{0}^{\infty}dL\left(\langle R_{12}|\mathcal{Q}^{\alpha}(L)\rangle_{1}|\pi_{\alpha}(L)\rangle_{2}+\langle R_{12}|\mathcal{T}^{\alpha}(L)\rangle_{1}[\hat{Q},|\pi_{\alpha}(L)\rangle_{2}]\right)\,,
\]
where 
\[
|\mathcal{Q}^{\alpha}(L)\rangle\equiv[\hat{Q},|\mathcal{T}^{\alpha}(L)\rangle]\,.
\]
Although $|\mathcal{T}^{\alpha}(L)\rangle$ and $|\mathcal{Q}^{\alpha}(L)\rangle$
satisfy
\begin{eqnarray}
 &  & \left[\lim_{\tau\to\infty}\langle\negthinspace\langle0|e^{-\tau\hat{H}}\right]|\mathcal{T}^{\alpha}(L)\rangle=0\,,\label{eq:Talpha}\\
 &  & \left[\lim_{\tau\to\infty}\langle\negthinspace\langle0|e^{-\tau\hat{H}}\right]|\mathcal{Q}^{\alpha}(L)\rangle=0\,,\label{eq:Qalpha}
\end{eqnarray}
$[\hat{Q},\hat{H}]$ does not vanish. 

Since we need to define the physical states by the BRST transformation,
we want to have a BRST invariant formulation. Such a formulation is
given by introducing auxiliary fields $|\lambda_{\alpha}^{\mathcal{T}}(\tau,L)\rangle,|\lambda_{\alpha}^{\mathcal{Q}}(\tau,L)\rangle$
and modifying the Euclidean action (\ref{eq:Eucaction}) as follows:
\begin{eqnarray*}
I_{\mathrm{BRST}} & = & \int_{0}^{\infty}d\tau\left[-\int_{0}^{\infty}dL\langle R_{12}|\pi_{\alpha}(\tau,L)\rangle_{1}\frac{\partial}{\partial\tau}|\phi^{\alpha}(\tau,L)\rangle_{2}+H(\tau)\right.\\
 &  & \hphantom{\int_{0}^{\infty}d\tau\quad}\left.+\int_{0}^{\infty}dL\left(\langle R_{12}|\mathcal{Q}^{\alpha}(\tau,L)\rangle_{1}|\lambda_{\alpha}^{\mathcal{Q}}(\tau,L)\rangle_{2}+\langle R_{12}|\mathcal{T}^{\alpha}(\tau,L)\rangle_{1}|\lambda_{\alpha}^{\mathcal{T}}(\tau,L)\rangle_{2}\right)\vphantom{\int_{0}^{\infty}dL\langle R|\pi_{\alpha}(\tau,L)\rangle\frac{\partial}{\partial\tau}|\phi^{\alpha}(\tau,L)\rangle}\right]\,.
\end{eqnarray*}
$I_{\mathrm{BRST}}$ is invariant under the transformation 
\begin{eqnarray}
\delta_{\eta}|\phi^{+}(\tau,L)\rangle & = & \eta P_{-}Q|\phi^{+}(\tau,L)\rangle\,,\nonumber \\
\delta_{\eta}|\phi^{-}(\tau,L)\rangle & = & \eta Q|\phi^{-}(\tau,L)\rangle-\eta b_{0}^{-}P\partial_{L}|\phi^{+}(\tau,L)\rangle\,,\nonumber \\
\delta_{\eta}|\pi_{+}(\tau,L)\rangle & = & \eta Q|\pi_{+}(\tau,L)\rangle-\eta b_{0}^{-}P\partial_{L}|\pi_{-}(\tau,L)\rangle\,,\nonumber \\
\delta_{\eta}|\pi_{-}(\tau,L)\rangle & = & \eta P_{-}Q|\pi_{-}(\tau,L)\rangle\,,\nonumber \\
\delta_{\eta}|\lambda_{\alpha}^{\mathcal{Q}}(\tau,L)\rangle & = & \eta\left[|\pi_{\alpha}(\tau,L)\rangle+|\lambda_{\alpha}^{\mathcal{T}}(\tau,L)\rangle\right]\,,\nonumber \\
\delta_{\eta}|\lambda_{\alpha}^{\mathcal{T}}(\tau,L)\rangle & = & -\delta_{\eta}|\pi_{\alpha}(\tau,L)\rangle\,.\label{eq:BRST}
\end{eqnarray}
The correlation functions defined by 
\begin{equation}
\frac{\int[d\pi d\phi d\lambda^{\mathcal{Q}}d\lambda^{\mathcal{T}}]e^{-I_{\mathrm{BRST}}}|\phi^{\alpha_{1}}(0,L_{1})\rangle\cdots|\phi^{\alpha_{n}}(0,L_{n})\rangle}{\int[d\pi d\phi d\lambda^{\mathcal{Q}}d\lambda^{\mathcal{T}}]e^{-I_{\mathrm{BRST}}}}\,.\label{eq:IBRSTcorr}
\end{equation}
can be proved to coincide with those in (\ref{eq:Icorr}) using (\ref{eq:Talpha})
and (\ref{eq:Qalpha}) \cite{Ishibashi2023}. The BRST transformation
(\ref{eq:BRST}) can be used to define the physical states. 

\section{Discussions\label{sec:Discussions}}

In this paper, we have constructed a string field theory based on
the Strebel differential and the combinatorial moduli space. The formulation
of the theory looks exactly like that of the theory \cite{Ishibashi2023}
based on the pants decomposition of hyperbolic surfaces. The intrinsic
reason for this is that the combinatorial moduli space arises \cite{mondello2011riemann,mondello2009triangulated,do2010asymptotic}
in the long boundary limit of the description using the hyperbolic
surfaces. Actually the recursion relations (\ref{eq:recursion}) and
(\ref{eq:recursion11}) can be obtained by taking the long string
limit of those for the hyperbolic case, assuming that the hyperbolic
amplitudes defined in \cite{Ishibashi2023} become the combinatorial
one defined in this paper in the limit. 

The propagator of the string field theory corresponds to a cylinder
of vanishing height as in the hyperbolic case. Such formulations are
pretty unconventional. It is usually assumed that the kinetic terms
of a string field theory should yield those for the elementary particles
contained in the theory. Moreover, in our theory, both the propagator
and vertices represent local interactions of strings and nonlocality
resides only in the propagation of the external strings, as was mentioned
in section \ref{sec:Strebel-differentials-and}. Such a description
looks very unphysical, but the formulation may be suitable for studying
the tensionless limit of string theory \cite{Gopakumar2005,Bhat2022}. 

The Fokker-Planck Hamiltonian and the Euclidean action we obtain consist
of kinetic terms and three string interaction terms. It will be an
interesting future problem to explore classical solutions of the
theory, in particular closed string tachyon solutions \cite{Okawa2004,Yang2005,Moeller2007,Scheinpflug2023}.
Another thing to do is to generalize the formulation to the superstring
case. In doing so, the recent results \cite{schwarz2023super} on
the combinatorial description of the supermoduli space may be helpful. 

\section*{Acknowledgments}

This work was supported in part by Grant-in-Aid for Scientific Research
(C) (18K03637) from MEXT.

\appendix

\section{Combinatorial moduli space\label{sec:Combinatorial-moduli-space}}

In this appendix, we will give a brief review of Strebel differentials
and combinatorial moduli space. We refer the reader to \cite{Gopakumar2005,Saadi1989,Zwiebach1991,mulase1998ribbon}
for details. 

\subsection{Strebel differentials}

Let us consider quadratic differentials 
\[
\varphi(z)dz^{2}\,,
\]
on a compact Riemann surface. With a nonzero quadratic differential
$\varphi(z)dz^{2}$, one can define a locally flat metric 
\begin{equation}
ds^{2}=\left|\varphi(z)\right|dzd\bar{z}\,,\label{eq:ds2}
\end{equation}
on the surface. A horizontal trajectory of the quadratic differential
$\varphi(z)dz^{2}$ is a curve along which $\varphi(z)dz^{2}$ is
real and positive. The horizontal trajectories are either closed or
nonclosed. Jenkins-Strebel quadratic differentials are those for which
the union of nonclosed horizontal trajectories covers a set of measure
zero. The nonclosed horizontal trajectories of a Jenkins-Strebel quadratic
differential decompose the surface into ring domains swept by closed
horizontal trajectories. These ring domains can be annuli or punctured
disks. %
\begin{comment}
A curve $z(t)\ (0<t<1)$ on $\mathrm{C}$ is called horizontal with
respect to $\varphi(z)$ if
\[
\varphi(z(t))\left(\frac{dz}{dt}\right)^{2}>0\,,
\]
holds for $0<t<1$. If the curve satisfies
\[
\varphi(z(t))\left(\frac{dz}{dt}\right)^{2}<0\,,
\]
instead, it is called vertical. 
\end{comment}

Let $\mathrm{C}$ be a compact Riemann surface of genus $g$ with
$n$ marked points ${p_{1},\cdots,p_{n}}$ with
\[
n\geq1,\ 2g-2+n>0\,.
\]
Strebel \cite{strebel1984quadratic} proved that for any such $\mathrm{C}$
and any $n$-tuple of positive numbers $(L_{1},\cdots,L_{n})$, there
exists a unique Jenkins-Strebel differential $\varphi(z)dz^{2}$ which
satisfies the following conditions: 
\begin{itemize}
\item $\varphi(z)dz^{2}$ has double poles at $p_{a}\ (a=1,\cdots,n)$ and
is holomorphic on $\mathrm{C}\backslash\left\{ p_{1},\cdots,p_{n}\right\} $. 
\item The maximal ring domains of $\varphi(z)dz^{2}$ are $n$ punctured
disks $D_{a}\ (a=1,\cdots,n)$ around $p_{a}$, which is swept out
by closed trajectories with length $L_{a}$ with respect to the metric
(\ref{eq:ds2}). 
\end{itemize}
This unique quadratic differential is called the Strebel differential. 

The nonclosed trajectories of the Strebel differential decompose $\mathrm{C}$
into disks $D_{a}$ around $p_{a}$. It is possible to choose a local
coordinate $u_{a}$ around $p_{a}$, so that the Strebel differential
$\varphi(z)dz^{2}$ is expressed as
\begin{equation}
\varphi(z)dz^{2}=-\left(\frac{L_{a}}{2\pi}\right)^{2}\frac{\left(du_{a}(z)\right)^{2}}{\left(u_{a}(z)\right)^{2}}\,,\label{eq:double}
\end{equation}
 $p_{a}$ corresponds to $u_{a}=0$ and $\partial D_a$ corresponds to $|u_a|=1$. $u_{a}(z)$ can be written
as 
\begin{equation}
u_{a}(z)=e^{\pm\frac{2\pi i}{L_{a}}\int_{z_{0}}^{z}dz^{\prime}\sqrt{\varphi(z^{\prime})}}\,.\label{eq:uaz}
\end{equation}
The metric on the disk $D_{a}$ becomes
\begin{equation}
ds^{2}=\left(\frac{L_{a}}{2\pi}\right)^{2}\frac{\left|du_{a}\right|^{2}}{\left|u_{a}\right|^{2}}\,.\label{eq:cylinder}
\end{equation}
Therefore the disk becomes a semi-infinite cylinder with circumference
$L_{a}$ and $p_{a}$ can be regarded as a puncture. 

\subsection{Ribbon graphs}

Given a Strebel differential, the union of the nonclosed trajectories
and the zeros is called its critical graph. A critical graph of a
Strebel differential becomes a so-called ribbon graph. 

A ribbon graph is a graph made from vertices and edges with a cyclic
orientation of the half edges meeting at each vertex. We restrict
ourselves to connected graphs such that every vertex has degree at
least three. The cyclic ordering allows the edges to be thickened
in a canonical way and we get a graph made of ribbons. For such a
graph, one can define its boundary. If a ribbon graph possesses $e$
edges, $v$ vertices and $n$ boundary components, the genus $g$
of the graph is defined to satisfy 
\[
v-e+n=2-2g\,.
\]
 The number of the boundary components of the critical graph of a
Strebel differential is equal to that of the punctures and the genus
$g$ is equal to that of the Riemann surface. A ribbon graph of genus
$g$ and with $n$ labeled boundary components is called a ribbon
graph of type $(g,n)$.

We can assign lengths to the edges of the critical graph. A ribbon
graph with lengths of the edges assigned is called a metric ribbon
graph. The set of equivalence classes of metric ribbon graphs of type
$(g,n)$ with respect to symmetry is denoted by $\mathcal{M}_{g,n}^{\mathrm{comb}}$,
which is called the combinatorial moduli space. $\mathcal{M}_{g,n}^{\mathrm{comb}}$
can be decomposed into cells such that a cell corresponds to a ribbon
graph $\Gamma$ of type $(g,n)$. The cell for $\Gamma$ is parametrized
by the lengths $l_{1},\cdots,l_{e(\Gamma)}$ of the edges of $\Gamma$,
where $e(\Gamma)$ denotes the number of edges of $\Gamma$. Therefore
$\mathcal{M}_{g,n}^{\mathrm{comb}}$ can be identified with
\begin{equation}
\bigcup_{\Gamma}\frac{\mathbb{R}_{+}^{e(\Gamma)}}{\mathrm{Aut}(\Gamma)}\,,\label{eq:cell}
\end{equation}
where $\mathrm{Aut}(\Gamma)$ denotes the automorphism group of $\Gamma$
preserving the boundary labeling. One can go to an adjacent cell by
taking the length of an edge to be $0$ or by expanding out a collapsed
vertex. 

From the theorem of Strebel, we can see that there is a map
\begin{equation}
\mathcal{M}_{g,n}\times\mathbb{R}_{+}^{n}\to\mathcal{M}_{g,n}^{\mathrm{comb}}\,,\label{eq:map}
\end{equation}
where $\mathcal{M}_{g,n}$ denotes the moduli space of Riemann surfaces
of genus $g$ with $n$ punctures and $\mathbb{R}_{+}^{n}$ represents
the parameters $(L_{1},\cdots,L_{n})$. Actually there exists the
inverse of the map (\ref{eq:map}) and $\mathcal{M}_{g,n}\times\mathbb{R}_{+}^{n}$
and $\mathcal{M}_{g,n}^{\mathrm{comb}}$ are homeomorphic.

For our purpose, it is convenient to define $\mathcal{M}_{g,n}^{\mathrm{comb}}(\mathbf{L})$
which is the set of equivalence classes of metric ribbon graphs of
genus $g$ with $n$ labeled boundary components whose lengths are
$\mathbf{L}=(L_{1},\cdots,L_{n})$. Locally $\mathcal{M}_{g,n}^{\mathrm{comb}}(\mathbf{L})$
is parametrized by the lengths of the edges of the metric ribbon graph
which satisfy the constraint that the lengths of the boundary components
are $\mathbf{L}=(L_{1},\cdots,L_{n})$. %
\begin{comment}
On $\mathcal{M}_{g,n}^{\mathrm{comb}}(\mathbf{L})$, Kontsevich defined
a symplectic form $\omega_{\mathrm{K}}$, which played an essential
role in proving the Witten's conjecture \cite{kontsevich1992intersection}.
\end{comment}
{} From a metric ribbon graph $G\in\mathcal{M}_{g,n}^{\mathrm{comb}}(\mathbf{L})$,
one can construct a punctured Riemann surface $\mathrm{gr}_{\infty}(G)\in\mathcal{M}_{g,n}$
by attaching semi-infinite cylinders to the boundary, with the metric
(\ref{eq:cylinder}) for the $a$-th cylinder. 
\[
\mathrm{gr}_{\infty}:\mathcal{M}_{g,n}^{\mathrm{comb}}(\mathbf{L})\to\mathcal{M}_{g,n}\,,
\]
is a homeomorphism which is real analytic on each cell \cite{mondello2011riemann}. 

\section{Non-admissible twists\label{sec:Non-admissible-twists}}

\begin{figure}
\begin{centering}
\includegraphics[scale=0.7]{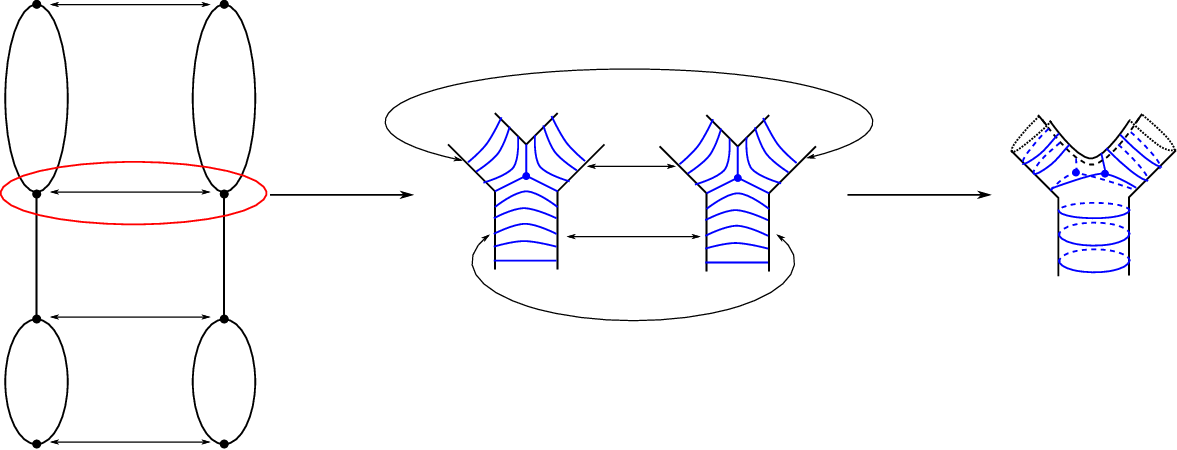}
\par\end{centering}
\caption{An example of non-admissible twist. \label{fig:An-example-of}}
\end{figure}

As is mentioned in section \ref{sec:Strebel-differentials-and},
we can decompose $\left|G_{g,n,\mathbf{L}}\right|$ into $2g-2+n$
pairs of pants by cutting it along $3g-3+n$ essential simple
closed curves. Each pair of pants inherits the measured foliation
which is equivalent to that of $\left|G_{0,3,(L_{1},L_{2},L_{3})}\right|$
for some $L_{1},L_{2},L_{3}$. On the other hand, if we glue surfaces
of the form $\left|G_{0,3,\cdot}\right|$ with the measured foliations
in a measure preserving way, we cannot get a surface of the form $\left|G_{g,n,\mathbf{L}}\right|$
for some choice of twists. Such a twist is called a non-admissible
twist. 

An example of such a twist is illustrated in Figure \ref{fig:An-example-of}.
Let us consider two dumbbell type combinatorial pairs of pants of
the same shape and glue them with the zero twist as shown in the figure.
The thickened pairs of pants are glued in the way so that the foliation
develops leaves connecting singular points\footnote{Such a leaf is called a saddle connection. }
and a part of the surface looks like a closed string propagator. Such
a foliation cannot come from a metric ribbon graph. 

\providecommand{\href}[2]{#2}\begingroup\raggedright\endgroup

\end{document}